\documentclass[fleqn, usenatbib]{mnras}

\usepackage{cancel}
\usepackage{newtxtext,newtxmath}
\usepackage[dvipsnames]{xcolor}
\usepackage{ae,aecompl}
\usepackage{natbib}
\usepackage{graphicx}	
\usepackage{amsmath}	
\usepackage{amssymb}	

\newcommand{\ie}{{\textit{i.e.},~}}
\newcommand{\eg}{{\textit{e.g.},~}}

\DeclareMathOperator{\Tr}{Tr}
\title[]{\centering A Novel CMB Component Separation Method:\\
Hierarchical Generalized Morphological Component Analysis}
\author[]{Sebastian Wagner-Carena$^{1,2}$\thanks{E-mail: swagnerc@stanford.edu}, Max Hopkins$^{3,4}$\thanks{E-mail: nmhopkin@eng.ucsd.edu}, Ana Diaz Rivero$^{1}$, 
Cora Dvorkin$^{1}$\thanks{E-mail: cdvorkin@g.harvard.edu}\\
\\
$^1$ Department of Physics, Harvard University, Cambridge, MA 02138, USA\\
$^2$ Department of Physics, Stanford University, Stanford, CA 94305, USA\\
$^3$ Department of Mathematics, Harvard University, Cambridge, MA 02138, USA\\
$^4$ Department of Computer Science and Engineering, University of California San Diego, San Diego, CA 92092, USA\\
}

\pubyear{2020}

\hypersetup{final}
\begin{document}
\label{firstpage}
\pagerange{\pageref{firstpage}--\pageref{lastpage}}
\maketitle

\begin{abstract}
We present a novel technique for Cosmic Microwave Background (CMB) foreground subtraction based on the framework of blind source separation. Inspired by previous work incorporating local variation to Generalized Morphological Component Analysis (GMCA), we introduce Hierarchical GMCA (HGMCA), a Bayesian hierarchical graphical model for source separation. We test our method on $N_{\rm side}=256$ simulated sky maps that include dust, synchrotron, free-free and anomalous microwave emission, and show that HGMCA reduces foreground contamination by $25\%$ over GMCA in both the regions included and excluded by the \textit{Planck} UT78 mask, decreases the error in the measurement of the CMB temperature power spectrum to the $0.02-0.03$\% level at $\ell>200$ (and $<0.26\%$ for all $\ell$), and reduces correlation to all the foregrounds. We find equivalent or improved performance when compared to state-of-the-art Internal Linear Combination (ILC)-type algorithms on these simulations, suggesting that HGMCA may be a competitive alternative to foreground separation techniques previously applied to observed CMB data. Additionally, we show that our performance does not suffer when we perturb model parameters or alter the CMB realization, which suggests that our algorithm generalizes well beyond our simplified simulations. Our results open a new avenue for constructing CMB maps through Bayesian hierarchical analysis.
\end{abstract}


\section{Introduction}
The Cosmic Microwave Background (CMB) is the earliest observable light in the universe and one of the most important observables for constraining models of early-universe physics. The Lambda Cold Dark Matter ($\Lambda$CDM) model, inflation, and large-scale structure can all be better understood by more precise measurements of the CMB, as can be seen from the range of scientific results obtained with data from the recent \textit{Planck} satellite \citep{planck2018overview}.
A wide range of applications require an accurate reconstruction of the CMB sky maps: extraction of the gravitational lensing and integrated Sachs-Wolfe (ISW) signals, constraints  on  isotropy, global geometry, topological defects, and searches for  primordial non-Gaussianity  \citep{planck2013isw, planck2018lensing, planck2018isotropy,  planck2018png, McEwen:2016aob, Saadeh:2016sak}. Unfortunately, several sources, some within our own galaxy, others distant celestial objects, emit radiation at the same frequencies as the CMB. Separating these foreground sources from the CMB has become a critical problem for modern cosmology. \\
\indent Traditionally, cosmologists have used template fitting techniques and second order statistical methods to isolate the CMB \citep{gorski1996power, hobson1998foreground,bedini2005separation,martinez2003cosmic,ILC,NILC,SILC,SEVEM,WMAP,tegmark2003high}.
However, recent work has suggested that approaching the problem from a Bayesian perspective could yield CMB maps that are more accurate \citep{SMICAorig,SMICAmid,wavelets,Bobin,nGMCA,L-GMCA}. Bayesian techniques relax assumptions about second or higher-order statistics, and can be adapted to leverage alternative properties of the CMB and foreground components. In this paper we focus on methods that exploit sparsity - the assumption that the CMB and foreground components have small support in a specific basis. Our particular basis of interest is constructed from scale-discretized wavelets - oscillating functions defined both in real and harmonic space that can be used to decompose signals on the sphere. Extensive work has been done on constructing and computing accurate wavelets on the sphere \citep{antoine1999wavelets,narcowich2006localized,wiaux2008exact}, and there are numerous examples of the use of wavelets in the context of CMB data \cite{wavelets, NILC, planck2013components, planck2015foregrounds,McEwen:2016aob,SILC, planck2018foregrounds}.

Adding a Bayesian prior on the sources that enforces sparsity is the cornerstone of the Generalized Morphological Component Analysis (GMCA) algorithm \citep{Bobin}, which effectively separates the CMB from foregrounds. An extension to GMCA, coined Local GMCA (LGMCA) \citep{L-GMCA}, attempts to better capture the local variance of the foreground sources, circumventing one of the main limitations of vanilla GMCA.
By running multiple independent GMCA models on different partitions of a sky map, the LGMCA method allows the foreground sources to display differing frequency dependence by position. In \cite{L-GMCA}, the authors showed that this method excels at producing full sky maps with minimal foreground, especially thermal Sunyaev-Z'eldovich (SZ) contaminants, but struggles with cosmic infrared background contamination at high multipole moments. While LGMCA improves on GMCA by incorporating local variation, it no longer falls into the Bayesian framework: there is no single probabilistic model that describes the data. Instead, LGMCA is the combination of many individual models that generate the data in different patches/at different scales and are therefore not necessarily consistent with one another\footnote{A detailed discussion of this property (and LGMCA) can be found in Section \ref{sec:LGMCA}.}. To reconstruct a final map, LGMCA is forced to use heuristic measurements to select from these reconstructions (for examples of possible heuristics see \citealt{L-GMCA}).

In this work we propose a new algorithm, Hierarchical Generalized Morphological Component Analysis (HGMCA), which captures the key improvement of LGMCA (incorporating local variation) in a single, formal generative probabilistic model. HGMCA runs GMCA across a number of partitions of the map while enforcing global consistency through the use of hierarchical priors. This enhances the reliability of the local fits and the ability to consider higher degrees of local variation compared to LGMCA. 

To highlight HGMCA's performance we compare it with GMCA and a wavelet-based internal linear combination (ILC) implementation on simulated CMB and foregrounds. We leave LGMCA out of the comparison because the code is not publicly available, and our implementation of LGMCA was not able to reproduce the improvements shown in \cite{L-GMCA}. The simulations are generated using the Python Sky Model (\texttt{PySM}) code \citep{PySM} at a resolution of $N_{\rm side} = 256$ and include synchrotron, thermal dust, anomalous microwave emission (AME), and free-free emission. These foreground emissions are modeled from \textit{Planck} templates and make conservative assumptions about their spectral dependence. While these simulations are of moderate resolution and realism, lacking differing beam resolutions and contaminants such as SZ clusters, they are sufficient to demonstrate the improvements resulting from the new algorithm: by using hierarchical priors, our model fits to local variations in the foreground frequency scaling without sacrificing the constraining power offered by the full map. This reduces the presence of foreground contamination in the reconstructed CMB signal as compared to GMCA and ILC-based methods (Section \ref{sec:results}).

In addition to showing HGMCA's reduction in foreground emissions in the reconstructed CMB, we also provide rigorous testing of its generalizability and robustness (Section \ref{sec:generalization}). We provide two types of testing for generalization: differing realizations of the CMB, and differing foreground templates. Since the hyperparameters of HGMCA are trained using \texttt{PySM} simulations, to ensure that we accurately represent the performance of HGMCA, we present all results across simulations with a different CMB realization than used in training. This includes our presented comparisons to GMCA and wavelet ILC (WavILC), an ILC implementation based on the work of \cite{planck2018foregrounds,NILC3,SILC}. To ensure that we have not overfit hyperparameters to specific foreground templates, we also show HGMCA's performance over simulations which use different foreground templates with smaller-scale local variation. Our results suggest that when the map is subdivided to a sufficient extent, HGMCA is robust both to changes in CMB and foreground realizations. Finally, we show HGMCA's robustness to changes in the hierarchical prior and sparsity hyperparameters by demonstrating that altering these parameters by up to 20\% in either direction has a negligible effect on performance.

In the following sections, we briefly review the existing foreground separation techniques (Section \ref{sec:br}), describe our model from a probabilistic perspective (Section \ref{sec:HGMCA}), explain the implementation of our optimization algorithm (Section \ref{sec:imp}), and compare the results of our method to those of GMCA and a state of the art ILC CMB reconstruction on a set of simulations (Section \ref{sec:results}). Finally, we show that these improvements do not appear to be sensitive to perturbations in our model parameters, the random seed, or the specifics of our simulation (Section \ref{sec:generalization}). 
We conclude in Section \ref{sec:conclusion}, and mention possible extensions and applications of our HGMCA method to future CMB experiments.

\section{Background}\label{sec:br}
Our work follows a long line of research in source separation for the CMB. All of the techniques that we discuss (and which are used by the \textit{Planck} collaboration, \citealt{planck2013cosmo, planck2015cosmo, planck2018cosmo})
view the sky as a linear combination of the CMB, foreground components, and instrumental noise. In matrix form, this can be written as 
\begin{equation}\label{eq:BSS}
X= A  S+  N,
\end{equation}
where $X$ is the observed data vector, $A$ is referred to as the mixing matrix, $S$ is the source vector, and $N$ is the noise. Each row of $X$ is a sky map at a different frequency, each row of $S$ corresponds to a different source -- for instance, independent components of signal in the sky, including both the CMB and sources of foreground -- and the coefficients $a_{ij}$ of the mixing matrix $A$ dictate how much source $j$ contributes at frequency $i$. The dimensions of $A$ and $S$ depend on the number of sources ($N_\mathrm{S}$).
 Note that assuming a single mixing matrix $A$ implies that the amount a source contributes at a given frequency does not change spatially, which is known to not hold true for the foreground sources \citep{dunkley2009five,fuskeland2014spatial,planck2015foregrounds}. Incorporating local frequency dependence requires a modification to the framework and presents a unique challenge: finding a source separation model which allows local variation while maintaining global consistency.

Many different techniques have been developed to separate the CMB from observed data based on Equation~\eqref{eq:BSS}. In Appendix~\ref{app:rw}, we review algorithms used by \textit{Planck} from the three main paradigms: template fitting, internal linear combination (ILC), and Bayesian inference. In the following sections we provide some background on the lattermost technique, and discuss GMCA and LGMCA -- the main predecessors to HGMCA.

\subsection{Bayesian Source Separation}

Bayesian methods aim to isolate not just the CMB, but examine the entire posterior distribution $P(A, S|X)$. When done in a blind manner (\ie with few assumptions about both $A$ and $S$), this reduces to a Bayesian version of a problem which has received wide attention in the machine learning and statistics communities known as Blind Source Separation (BSS). Popular methods for examining the posterior include solving for the maximum a posteriori estimate, including variants of independent component analysis (ICA) and GMCA \citep{ICA,SMICAorig,Bobin,L-GMCA}, or estimation via sampling \citep{geman1984stochastic}.
Bayesian techniques such as these also allow for sophisticated priors on $A$ and $S$ such as non-negativity (a problem referred to as non-negative matrix factorization) \citep{nGMCA}, or compact and simple representation in particular bases, like wavelets \citep{Bobin}. The widespread ILC-type algorithms discussed in Appendix \ref{app:ILC} do not have analytic solutions that satisfy the constraints provided by such priors; therefore, Bayesian techniques are better suited to take advantage of known properties of the CMB and foregrounds. 

\subsection{GMCA} \label{sec:GMCA}
While previous source separation techniques, such as wavelet SMICA \citep{wavelets}, had already exploited the distinction of the CMB and foreground sources in wavelet bases, none had used the fact that these sources are also sparse. \cite{Bobin} introduced Generalized Morphological Component Analysis (GMCA) to leverage the sparsity of the distinct signals as an additional constraint. GMCA begins by assuming that the noise $N$ in the BSS model is Gaussian, such that we can rewrite the model as a conditional probability distribution\footnote{Throughout this work, when we write $X \sim \mathcal{N}(A S,\sigma_{X} I)$ we are implicitly invoking a vector notation of the form $\vec{X} \sim \mathcal{N}(\vec{A S},\sigma_{X} I)$ where $\vec{X}$ is the vectorized form of $X$ and $\sigma_{X} I$ encodes the covariance of $\vec{A S}$.}:
\begin{equation}
X \sim \mathcal{N}(A S,\sigma_{X} I),
\end{equation}
where $I$ is the identity matrix and $\mathcal{N}$ is a multivariate Normal distribution. Here the covariance matrix $\sigma_{X} I$ is determined by the level of instrumental noise. Enforcing the sparsity of $S$ amounts to adding a leptokurtic prior to $S$ of the form \citep{Bobin}:
\begin{equation}
P(S) \ \varpropto \prod\limits_{i,j}e^{-\lambda |s_{i,j}|^{\gamma}}, \gamma \leq 1.
\end{equation}
As $\gamma$ goes to $0$, this returns a negative-log-likelihood (up to  an additive constant) of $\lambda||S||_0$, where $||S||_0$ counts the number of non-zero values of $S$ and is known as the $0$-norm.

To solve for $A$ and $S$, the authors compute its Maximum a Posteriori (MAP) estimates, which in the limit of small $\gamma$ amounts to solving the following optimization problem:
\begin{equation}\label{GMCA:argmin}
\text{argmin}_{A,S} \ \left(||X - AS||_F^2 + \lambda_S||S||_0 \right),
\end{equation}
where $S$ and $X$ have been transformed into the wavelet domain\footnote{Note that this conserves the linearity of the product $AS$.}, a convention that will be used throughout the paper unless otherwise noted. Appendix \ref{app:wavelet} gives a summary of the wavelet transform on the sphere, although the needlet wavelets used by the authors (see \citealt{bobin2014joint}) differ from the scale-discretized wavelets used in this work. Note that $||M||_F \equiv \sqrt{\Tr(M^TM)}$ is the Frobenius norm of a matrix $M$, and $\lambda_S$ is determined by the relative strength of the constant $\lambda$ in $P(S)$ along with the covariance $\sigma_{X} I$ of $P(X)$. 

Optimization for Equation~\eqref{GMCA:argmin} comes with a few inherent difficulties. First, the linear mixture model $X=AS$ has  a scaling degeneracy--if $A$ and $S$ are inversely scaled by a constant, their product remains invariant. To circumvent this issue, \cite{Bobin} assume that the columns of $A$ have a $2$-norm of $1$, and enforce this in their algorithm. We will make the same assumption in our optimization for HGMCA (see Appendix~\ref{app:HGMCA}), along with introducing a Bayesian interpretation of the assumption in Section~\ref{sec:HGMCA}.

Because $||S||_0$ is a combinatorial object, this optimization problem is, in general, NP-Hard, and thus intractable \citep{Bobin}. However, in many cases it is possible to either approximate or exactly solve the minimization problem through a convex relaxation--replacing the $0$-norm with a $1$-norm ($||S||_1=\sum\limits_{i,j}|S_{i,j}|$). The problem is then tackled via optimization techniques from compressed sensing such as LASSO shooting \citep{LASSO}, or, more generally, optimization techniques involving soft or hard thresholding. Another option is to directly approximate the $0$-norm solution through k-thresholding\footnote{Here we are referring to hard thresholding optimization schemes where all but k of the entries of a source are zeroed out.}. We will outline the thresholding and LASSO strategies in Section \ref{sec:imp_gmca}. For a more detailed discussion of these techniques in the context of GMCA, see \citep{Bobin}.

\begin{figure}
\centering
\includegraphics[scale=.8]{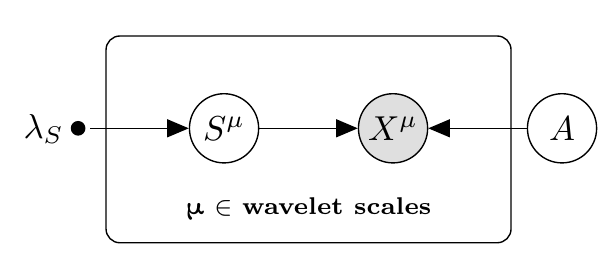}
\caption{Graphical model for the base GMCA algorithm. Variables are denoted by un-shaded circles, constants by dots, and observed data by shaded circles. Because our graphical model contains repeated sub-graphs, we use plate notation. A plate denotes a sub-graph repeated several times in the diagram, once per index provided in the plate label. Here, the plate stands for as many sub-graphs as there are wavelet scales $\mu$. Arrows, which indicate probabilistic dependence between variables, cross plate lines and are assumed to be repeated once per index. For example, here each observed $X^\mu$ is generated by a corresponding $S^\mu$ and the global mixing matrix $A$ (which is the same for all wavelet scales).}
\label{model:GMCA}
\end{figure}

As a visual aid, Figure \ref{model:GMCA} depicts the graphical model associated to the GMCA algorithm.

\subsection{LGMCA}\label{sec:LGMCA}

\begin{figure}
\centering
\includegraphics[scale=.3, trim = 0cm 5.8cm 0cm 0cm, clip]{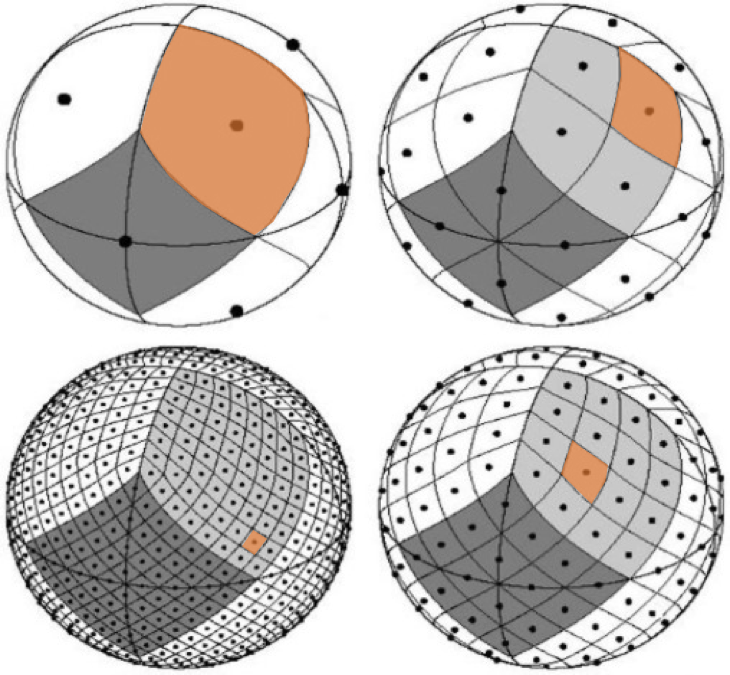}
\includegraphics[scale=.3, trim = 6cm 0cm 0cm 5.8cm, clip]{healpix_decomp.png}
\includegraphics[scale=.3, trim = 0cm 0cm 6cm 5.8cm, clip]{healpix_decomp.png}
\caption{HEALPix decomposition of sky map used by LGMCA at levels $l$ = 1,2,3, and 4, from left to right. Orange regions highlight the size of the patches of sky on which mixing matrices are calculated at a given level. Original HEALPix image credit: \protect\cite{Gorski:2004by}. }
\label{quad-tree}
\end{figure}

GMCA assumes that the mixing matrix is the same at every pixel, an assumption that does not hold for the heterogeneous emission in the sky, but is made for good reason: letting the mixing matrix vary across pixels is computationally intractable for problems such as the CMB. LGMCA \citep{L-GMCA}, the ``local'' modification of GMCA, incorporates locally-variable mixing matrices in a tractable manner, by exploiting a quad-tree decomposition to compromise between local adaptability and global consistency. This decomposition conveniently uses the hierarchical nature of the HEALPix pixelization scheme \citep{Gorski:2004by}.
Figure \ref{quad-tree} describes how this decomposition works on a full HEALPix map. At a given level $l$, the full map is decomposed into $4^l$ patches, meaning that, as $l$ increases, the patches probe smaller areas of the map. \footnote{This decomposition is done in the wavelet space (see Appendix \ref{app:wavelet}), meaning that the patches contain wavelet coefficients.}

In LGMCA, each level $l$ is treated as an independent probabilistic model, where GMCA is applied to each patch. Because each level is treated independently, the algorithm produces multiple reconstructions for the same patch of data. To merge these reconstructions into a final map, LGMCA dynamically chooses the level at which to reconstruct each patch of the sky by selecting the reconstruction with the lowest variance. \cite{L-GMCA} argue that this heuristic will select the level with the least foreground contamination due to the statistical independence of the CMB from the instrument noise and the foreground components.

The local variation permitted by LGMCA has been shown to reduce galactic foreground contamination overall compared to GMCA and ILC-type methods, with a particularly large improvement for Sunyaev-Zel'dovich contamination \citep{L-GMCA}. However, a drawback of LGMCA is that it does not admit a unified probabilistic structure. To see why, consider the simple case of a set of data divided into four equal patches. LGMCA could propose the global mixing matrix solution (i.e. the mixing matrix acquired by fitting to all four patches at once) for three of the patches and the local mixing matrix solution (i.e. the mixing matrix acquired from fitting to each patch separately) for the remaining patch. 
The issue with a probabilistic interpretation for LGMCA lies with this final patch, which must now show up twice in a probabilistic model - once associated to the global mixing matrix and once to its own local mixing matrix. Because LGMCA treats the generation of the data by the global and local mixing matrices separately, the same data are generated by two different, independent processes, which is inconsistent with a probabilistic model. Additionally, because deeper (higher $l$) levels have no dependence on their predecessors, small patches are ignorant of the constraining power offered by the rest of the sky and are thus prone to overfitting.

\section{Hierarchical GMCA}\label{sec:HGMCA}

\begin{figure}
\centering
\includegraphics[scale=.73]{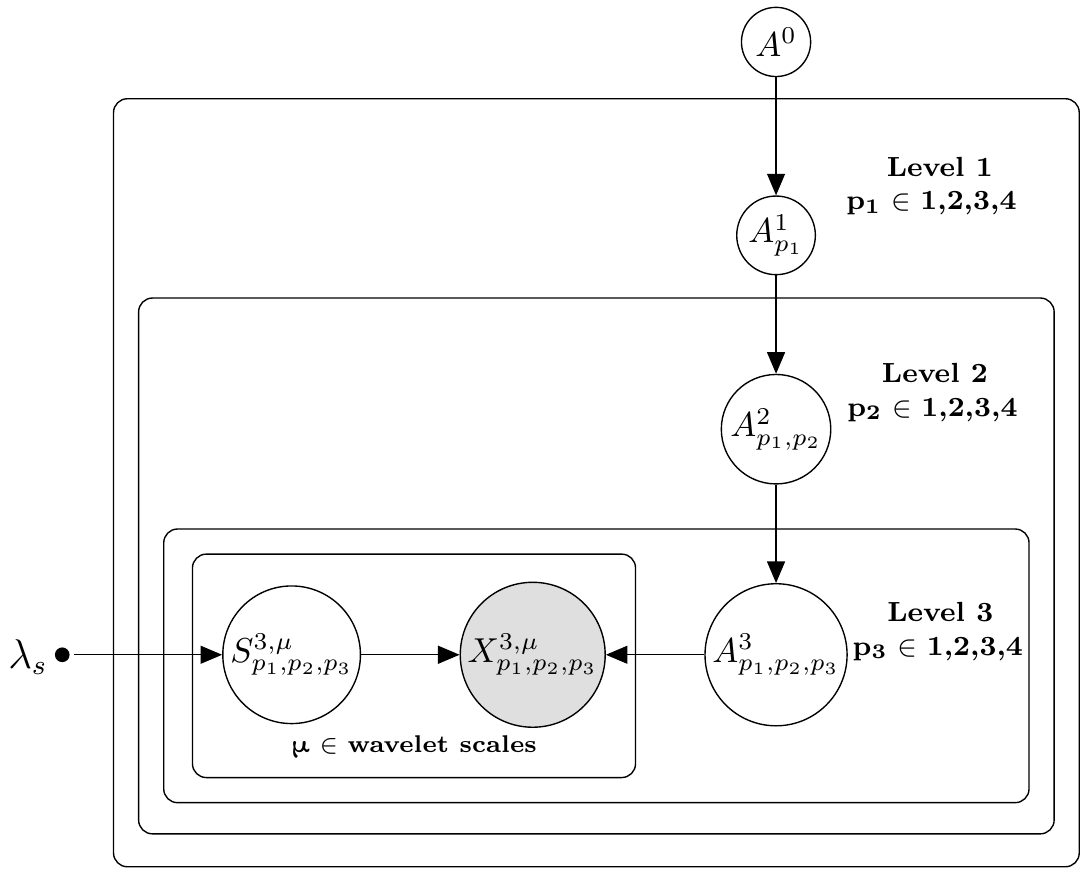}
\caption{Graphical model for Hierarchical GMCA at subdivision level $l=3$. $X^{3,\mu}_{p_1,p_2,p_3}$ represents the input maps at wavelet scale $\mu$ and patch $p_1,p_2,p_3$ of subdivision level $3$. $S^{3,\mu}_{p_1,p_2,p_3}$ represents the source at the same patch and wavelet scale, and $A^{3}_{p_1,p_2,p_3}$ represents the mixing matrix for that patch. The hierarchical nature of the mixing matrices is akin to a Hierarchical Bayesian Model \citep{gelman2013bayesian}, but the addition of the source matrix $S$ distances our model from this framework. For details on the patch notation see Section \ref{sec:HGMCA}. See Figure \ref{model:GMCA} for a description of the probabilistic graphical model notation.}
\label{model:simple}
\end{figure}

\begin{figure}
\centering
\includegraphics[scale=.6]{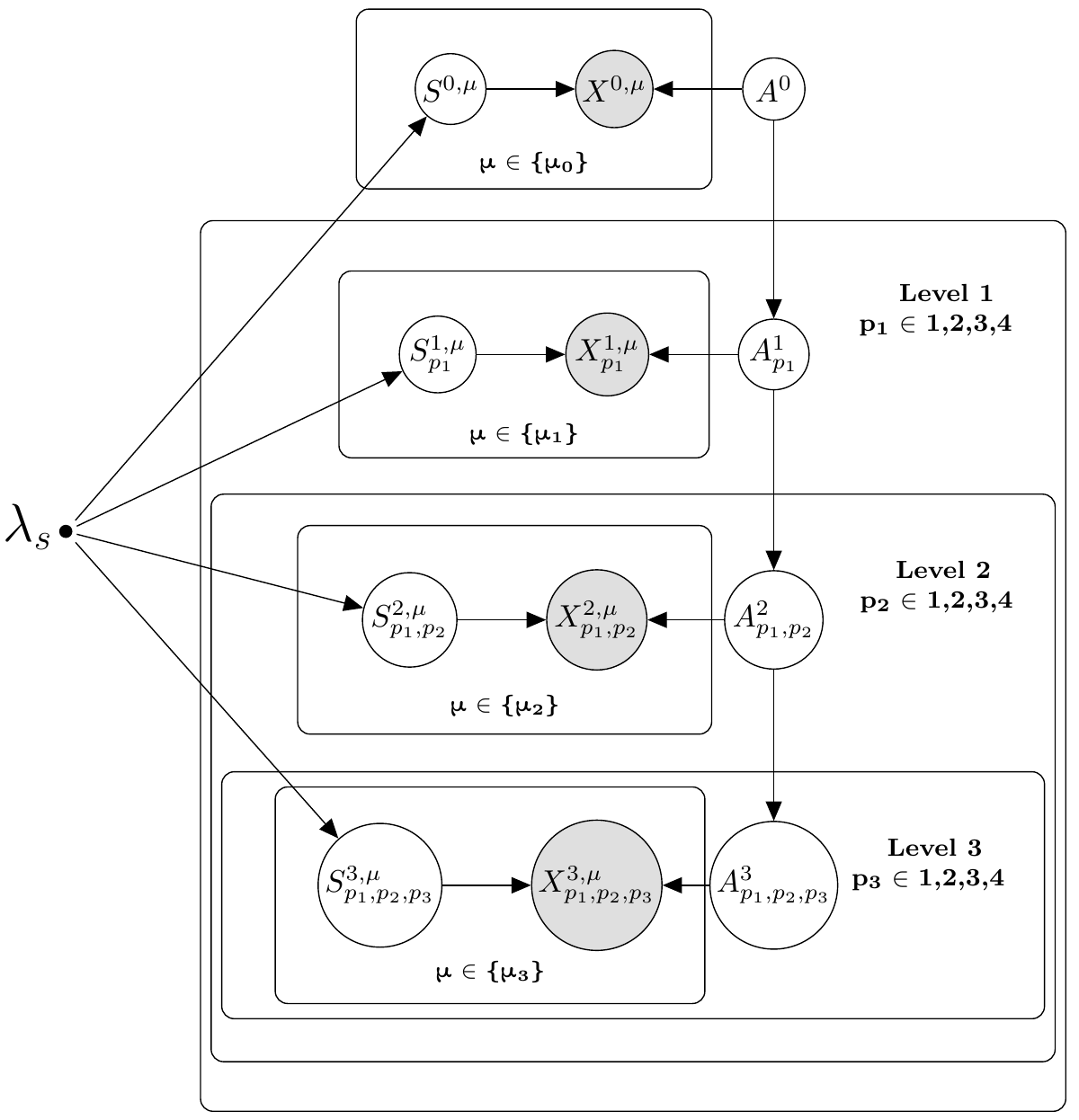}
\caption{Graphical model for Hierarchical GMCA at subdivision level $l=3$. Note that, unlike Figure \ref{model:simple}, this version of the model accounts for the resolution limit of each wavelet scale. This requires analyzing different wavelet scales at different levels of subdivision. $X^{l,\mu}_{p_1,\ldots p_l}$ represents the input maps at subdivision level $l$, wavelet scale $\mu$, and patch $p_1,\ldots p_l$ of subdivision level $l$. $S^{l,\mu}_{p_1,\ldots p_l}$ represents the source at the same patch and wavelet scale, and $A^{l}_{p_1,\ldots p_l}$ represents the mixing matrix for patch $p_1,\ldots p_l$. $\{\mu_l\}$ is the set of scales whose resolution limits them to being analyzed at level $l$. There is no overlap between $\{\mu_i\}$ and $\{\mu_j\}$ for 
$i \neq j$. See Figure \ref{model:GMCA} for a description of the probabilistic graphical model notation.}
\label{model:HGMCA}
\end{figure}

Rather than heuristically choosing between multiple independent models run at each level, we propose expanding LGMCA's notion of a quadtree decomposition to a single Bayesian hierarchical graphical model. Similar to GMCA, HGMCA aims to provide the MAP estimate of a posterior distribution. Unlike GMCA, this posterior distribution is spread across a hierarchy of mixing and source matrices, representing the entire probabilistic structure of HGMCA at subdivison level $l_{\rm max}$ (denoted HGMCA-$l_{\rm max}$):
\begin{align}\label{eqn:HGMCA}
    &P\bigl(\{A_p^{l}\},\{S_p^{\mu}\} | X \bigr) = \\
   & \quad \underbrace{\left(\prod_{\mu}P\bigl(S_p^{\mu}\bigr)\prod_l^{l_{\rm max}} \prod_{p \in \{p_l\}}  \mathcal{N}\bigl(X_p^{l,\mu} |A_p^{l}S_p^{\mu},\sigma_X I \bigr) P\bigl(A^{l}_{p}\bigr)\right)}_{\text{likelihood and priors}} \nonumber\\
   & \quad \times \quad \underbrace{\left(\prod_{l>0}^{l_{\rm max}} \prod_{p\in\{p_{l-1}\}} \prod_{p' \in [1,4]} \mathcal{N_{S,+}}\bigl(A_{p,p'}^{l}|A_{p}^{l-1},\sigma_A I \bigr)\right).}_{\text{Mixing matrix dependence across levels}.} \nonumber
\end{align}
Figure~\ref{model:simple} shows the corresponding graphical model to this posterior distribution for $l_{\rm max} = 3$ (HGMCA-3). Note that both Figure~\ref{model:simple} and Equation~\ref{eqn:HGMCA} ignore some complexity related to the support of the wavelet scales that we will introduce in Figure~\ref{model:HGMCA} and Equation \eqref{eq:loss}. Here the patch index $p$ at level $l$ corresponds to a tuple $(p_1,\ldots,p_l)$ and $\{p_l\}$ is the set of all such tuples at level $l$. The tuple notation corresponds to the structure of the hierarchy: for an $l-1$ tuple $p$, the level $l$ mixing matrix $A^l_{p,p'}$ is drawn from $A_p^{l-1}$ with $p' \in [1,4]$. $X_p^{l,\mu}$ is a sub-matrix of $X$ corresponding to the patch $p$ at level $l$ and wavelet scale $\mu$, and $A_p^{l}$ and $S_p^{l,\mu}$ are likewise the patch-specific mixing and source matrices. Note that the mixing matrix should not vary across different wavelet scales, and therefore there is only one mixing matrix per patch for all wavelet scales $\mu$. $X^{3,\mu}_{p_1,p_2,p_3}$, which is our observed variable, is generated from $A^3_{p_1,p_2,p_3}$ and $S^{3,\mu}_{p_1,p_2,p_3}$. The value of $P(S_p^{l,\mu})$ enforces the sparsity of the coefficients, with the exact distribution discussed in Section \ref{sec:GMCA}. $P(A^l_p)$ is a Heaviside function prior that enforces the non-negativity constraints discussed in Appendix \ref{ngmca} (in short, $A$ cannot be negative because it quantifies the contribution of a given source at a given frequency, which is positive for the CMB and its foregrounds\footnote{This does not hold true for SZ clusters, but one could easily remove the non negativity prior for a single source and add an SZ frequency scaling prior.}). 

We have made the assumption that a mixing matrix is drawn from a Gaussian defined by its prior and a covariance matrix $\sigma_A I$. More flexible choices can be made, but they would introduce additional parameters to the algorithm. The remainder of this section will be spent detailing the hierarchical dependence represented by Equation \eqref{eqn:HGMCA} and the optimization strategy for HGMCA.

Figure \ref{model:simple} shows the basic plate diagram describing the generative probabilistic structure of our model. In this example, we assume that the map is divided up to level $l=3$, which means that it is being divided into $4^3$ patches\footnote{In practice, HEALPix requires that the first level be divided into twelve rather than four patches. For notational simplicity, we will use four throughout the paper.}. At the level $l=0$, there is a mixing matrix $A^0$ that encodes the most general mixing matrix from which all the local mixing matrices are thenceforth drawn. While this mixing matrix does not directly generate the data, it is useful to think of it as the global mixing matrix for the full map. From this mixing matrix, a set of four additional mixing matrices $A^1_{p_1}$ are drawn from a Gaussian restricted to the positive orthant of the unit sphere, denoted $\mathcal{N_{S,+}}$\footnote{In more detail, each column is being drawn from a Von Mises-Fisher distribution restricted to the positive orthant.: $A^1_{p_1} \sim \mathcal{N_{S,+}}(A^0,\sigma_A I)$}. Each $A^1_{p_1}$ corresponds to one of the four subdivisions of the full map at level 1, $p_1 \in [1,4]$. If we were only running our analysis up to level 1, these mixing matrices would then be responsible for generating the data. However, in this example we go up to level $l=3$, therefore each patch $p_1$ is further subdivided into $4$ patches indexed by $p_2$, giving the corresponding mixing matrices $A^2_{p_1,p_2}\sim\mathcal{N_{S,+}}(A^1_{p_1},\sigma_A I )$. The process is repeated one final time to get the set of $64$ mixing matrices $A^3_{p_1,p_2,p_3}\sim\mathcal{N_{S,+}}(A^2_{p_1,p_2},\sigma_A I)$ that each correspond to one patch of the map at subdivision level $l=3$.

By constructing this hierarchical structure we can better encode and quantify the trade-off between local adaptability and global consistency. Each patch is constrained by the data on the rest of the map through the tree of priors, and the strength of this constraint is dictated by the ratio of $\sigma_X$ and $\sigma_A$. 

So far, we have assumed that we can subdivide the map at all scales and frequencies down to subdivision level 3. However, the support of the wavelets being used at scale $\mu$ effectively limits the resolution of the map $X^\mu$: the level $l$ of analysis used for $X^\mu$ cannot subdivide the map into patches that are smaller than that resolution limit.

To account for this, we modify the model such that the data at wavelet scale $X^\mu$ is analyzed at the highest $l$ permitted by the scale's resolution limit (recall that patch size decreases with increasing $l$). We call the set of scales whose resolution limits them to level $l$ $\{\mu_l\}$. Note that there is no overlap between $\{\mu_0\},\{\mu_1\},\{\mu_2\},$ and $\{\mu_3\},$ as each map at each scale can only be allowed to contribute once in the graphical model. Further, it is permissible that a set $\{\mu_{l'}\}$ be empty. A visual representation of HGMCA incorporating this additional source of complexity is shown in Figure \ref{model:HGMCA}.

Note that the mixing matrix does not vary within $\{\mu_l\}$, and therefore lacks the superscript $\mu$. On the other hand, we do not constrain the mixing matrix $A^{2}_{p_1,p_2}$ of patch $(p_1,p_2)$ to be the same as its parent $A^{1}_{p_1}$. This accounts for the higher order corrections to the local variation of the mixing matrix that scales at level $l=2$ are sensitive to but scales at $l=1$ are not.

\begin{figure*}
    \centering
    \includegraphics[scale=0.35]{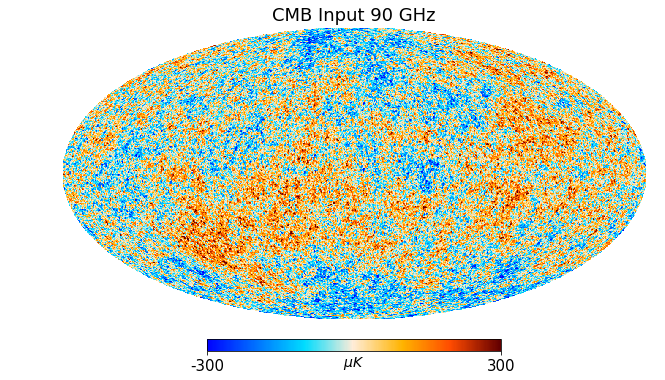}
    \includegraphics[scale=0.35]{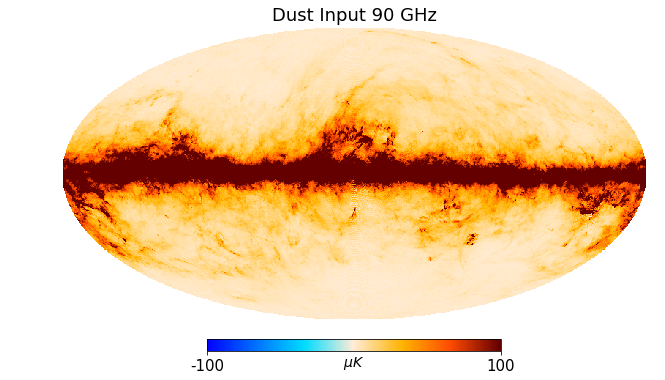}
    \includegraphics[scale=0.35]{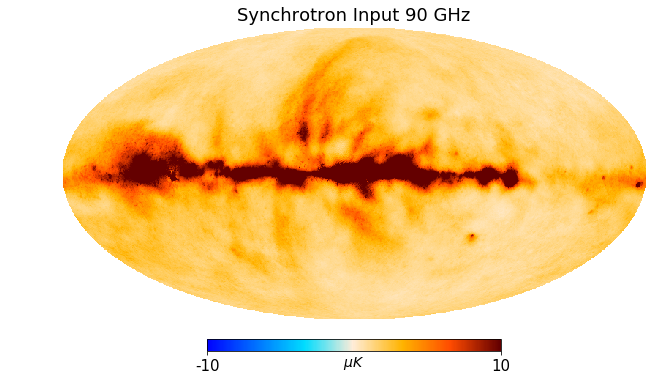}
    \includegraphics[scale=0.35]{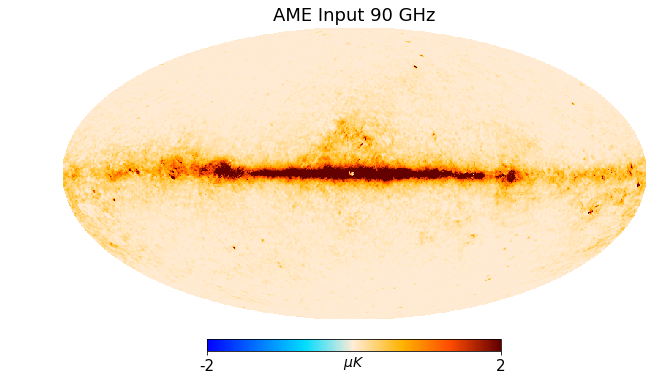}
    \includegraphics[scale=0.35]{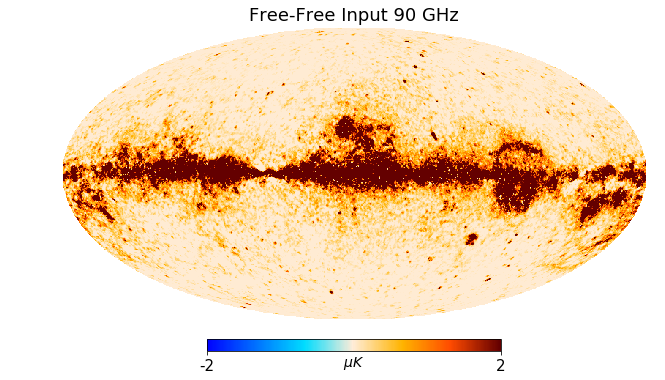}
    \includegraphics[scale=0.35]{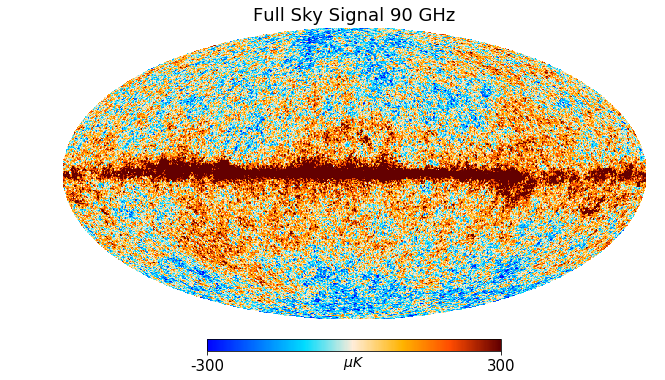}
    \caption{From top left to bottom right: the CMB, dust, synchrotron, anomalous microwave emission (AME), and free-free input maps that were combined to create the mock data used in this work. The full map with all signals combined is in the bottom right. Note that each input is presented at 90 GHz and has a different scaling for the color bar so that the structure is clearly visible.}
    \label{fig:inputs}
\end{figure*}


\begin{figure}
    \centering
    \includegraphics[scale=0.35]{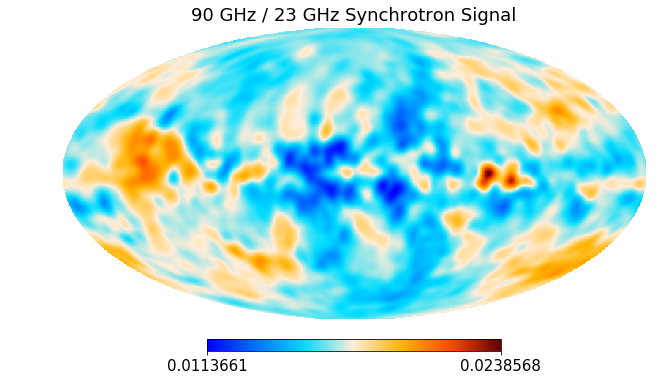}
    \includegraphics[scale=0.35]{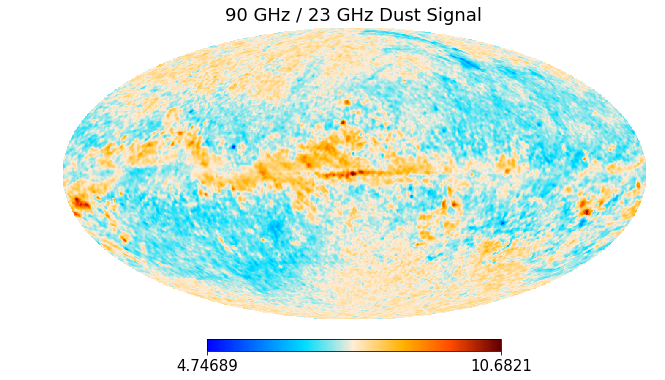}
    \caption{\textit{Top}: the ratio of the 90 GHz map to the 23 GHz map for a realization of synchrotron radiation. \textit{Bottom}: the same ratio for a realization of dust radiation. Both show local variation: without it, we would expect the ratios to return a constant value across the sky, but in these maps it varies spatially.}
    \label{fig:PySM}
\end{figure}

\section{Implementation}\label{sec:imp}

\subsection{GMCA Optimization}\label{sec:imp_gmca}
Computing the maximum a posteriori (MAP) of HGMCA reduces to running GMCA in each patch with additional priors from the mixing matrix hierarchy. \cite{Bobin} computes the MLE of GMCA via coordinate descent. By fixing all but a single column of the mixing matrix $A$ or row of the source matrix $S$, one can derive a closed form update that minimizes the negative log-posterior of the model. The results we present in this work are based on our own GMCA optimization strategy. While the mixing matrix optimization remains the same, we relax the 0-norm sparsity prior used in GMCA to a 1-norm, as suggested in \cite{Bobin} and discussed in Section \ref{sec:GMCA}. This allows us to write the closed form solution for a single row of our source matrix:
\begin{align}\label{eq:1norm}
    s_i &= (a^i)^T R_i -  \frac{\partial}{\partial s_i} \frac{\lambda_S}{2} ||s_i||_1,
\end{align}
and a single column of our mixing matrix:
\begin{align}\label{eq:Fast-gmca-a-sol}
a^i &= \left [\frac{R_is_i^T}{||R_is_i^T||_{F}} \right ]_+.
\end{align}
Here $s_i$ is a row of the source matrix, $a^{i}$ is the corresponding column of the mixing matrix, $R_i = X^l_p - \sum_{j\neq i} a^js_j$, and $[\ldots]_+$ indicates a non-negativity constraint enforced by setting negative values to 0. See Appendix \ref{app:HGMCA} and \cite{Bobin} for a more detailed derivation of these equations. 

Note that Equation \eqref{eq:1norm} contains a derivative with respect to the 1-norm $||s_i||_1$. There are a number of approximate methods to solve this type of optimization problem in the literature. Due to its relative simplicity of implementation, we use the LASSO shooting technique \citep{LASSO}, which gives the update equation:
\begin{align}
s_i &= \begin{cases}
(a^i)^T R_i - \frac{\lambda_S}{2} \text{sign} ((a^i)^T R_i) & |(a^i)^T R_i|> \frac{\lambda_S}{2} \\
0 & |(a^i)^T R_i| \leq \frac{\lambda_S}{2}
\end{cases}.
\end{align}
While standard LASSO shooting would suggest repeatedly conducting this update step until convergence, this is wasteful in the case of (H)GMCA as updating the remaining sources moves the point of convergence. Instead this update is only applied once before moving to the next source, and convergence is reached over many iterations over the rows of the source matrix and columns of the mixing matrix. We also empirically find that updating $S$ with the least squares solution (equivalent to setting $\lambda_s = 0$ in the loss function) every \textasciitilde 100 steps improves the rate of convergence and mitigates issues with local minima (tests of convergence are discussed in more detail in Section \ref{sec:conv}). We will call this version of the algorithm GMCA\textsuperscript{LASSO}.

A public version of the GMCA implementation can be found \href{https://github.com/jbobin/pyGMCALab/tree/master/pyGMCA}{online}\footnote{https://github.com/jbobin/pyGMCALab/tree/master/pyGMCA}. This version of GMCA implements an optimization procedure in line with what is described in \cite{Bobin}: it utilizes soft or hard median absolute deviation (MAD) thresholding\footnote{\cite{Bobin} suggest that while soft thresholding is more rigorous, the code returns better results with hard thresholding} to optimize the sparsity term and implements an alternating least squares version of GMCA known as "Fast GMCA." By assuming that the mixing matrix has nearly orthogonal columns, the update rule for this optimization only needs to fix one matrix at a time. Fixing $S$, the solution for the mixing matrix becomes
\begin{align}\label{eq:lstsq-A}
    A &= X S^T (S S^T)^{-1}.
\end{align}
When used for CMB estimation, \cite{Bobin} suggest fixing one column of the mixing matrix to the theoretical frequency dependence for the CMB, which we denote as $\tilde{a}^{CMB}$. This involves resetting that column back to $\tilde{a}^{CMB}$ after applying Equation~\eqref{eq:lstsq-A}.
Fixing $A$, the solution for the source matrix in the case of soft thresholding is:
\begin{align}\label{eq:s_sol}
s_i &= \begin{cases}
\left[ A_\text{pinv} X \right]_i  - \lambda_\text{MAD}(t) \text{sign} (\left[ A_\text{pinv} X \right]_i) & |\left[ A_\text{pinv} X \right]_i|>\lambda_\text{MAD}(t) \\
0 & |\left[ A_\text{pinv} X \right]_i| \leq \lambda_\text{MAD}(t)
\end{cases}.
\end{align}
where $\lambda_\text{MAD}(t)$ is a thresholding parameter scaled by the MAD of $\left[ A_\text{pinv} X \right]_i$ at iteration $t$. $\left[M\right]_i$ refers to the $i^\text{th}$ row of matrix $M$. Note that, unlike in GMCA\textsuperscript{LASSO}, $\lambda_\text{MAD}(t)$ is allowed to decrease with increasing iterations. The solution for hard thresholding does not subtract from the surviving source components:
\begin{align}
s_i &= \begin{cases}
\left[ A_\text{pinv} X \right]_i  & |\left[ A_\text{pinv} X \right]_i|>\lambda_\text{MAD}(t) \\
0 & |\left[ A_\text{pinv} X \right]_i| \leq \lambda_\text{MAD}(t)
\end{cases}.
\end{align}
In Appendix \ref{app:GMCA_Comparison} we compare Fast GMCA to our GMCA\textsuperscript{LASSO} implementation. While the optimum performance of both approaches is comparable, we find that GMCA\textsuperscript{LASSO} is significantly more stable to perturbations in the random seed and CMB realization. Since the main purpose of this work is to present our hierarchical approach, and we intend to compare the generalizability and performance of HGMCA to a wavelet-based ILC-type algorithm, we will default to the more stable GMCA\textsuperscript{LASSO}.

\subsection{HGMCA Optimization}
To compute the MAP estimate for HGMCA, we adopt a similar strategy. We first convert the posterior in Equation \eqref{eqn:HGMCA} into a loss function defined by the negative log-likelihood (with additive constants removed). We can write this as
\begin{align}\label{eq:loss}
    \mathcal{L} &= \underbrace{\sum_l  \sum_{\mu_l \in \{\mu_l\}} \sum_{p \in \{p_l\}}i^+_{\mathcal{S}}(A_p^l) + \lambda_S ||S_{p}^{l,\mu_l}||_1 
    + ||X_{p}^{l,\mu_l} 
    -A_{p}^{l}S_{p}^{l,\mu_l}||^2_F}_{\text{non-negativity prior, source prior, and reconstruction error}} \nonumber\\
    &+ \underbrace{\sum_{l>0} \sum_{p\in\{p_{l-1}\}} \sum_{i \in 1, \dots, 4}
    \lambda_A ||A_{p}^{l}-A_{p,i}^{l}||^2_F.}_{\text{A matrix dependence across levels},}
\end{align}
where $i^+_{\mathcal{S}}(A)$ is the characteristic function of positive matrices with normalized columns:
\begin{equation}
i^+_{\mathcal{S}}(A)=
\begin{cases}
0 & A \geq 0 \ \text{and} \ ||A^i||_2 = 1\\
\infty & \ \text{else.}
\end{cases}
\end{equation}
The factors $\lambda_A$ and $\lambda_S$ come from multiplying the log-likelihood by $\sigma_X^2$ and are ratios of the uncertainty in the draws of $X_{p}^{l,\mu_l}$ to the uncertainty in our priors. For example, $\lambda_A = \sigma_X^2/\sigma_A^2$.

As in the case of GMCA, while no tractable closed form solution for the mixing and source matrices in the loss function exists, it is possible to optimize by iterating over an approximate closed form\footnote{Stochastic gradient descent would be another possibility, but it performed poorly in GMCA\textsuperscript{LASSO}}. We use a coordinate descent strategy that is executed as follows. First, we initialize the sources and mixing matrices at all levels of subdivision. We find that using a short run of GMCA\textsuperscript{LASSO} as an initialization point rather than a random initialization improves convergence. We then extract an approximate closed form solution by fixing all but one row of the source matrix or column of the mixing matrix and minimizing the loss function. Here we state these solutions; their full derivation can be found in Appendix \ref{app:HGMCA}. 

For a given row of the source matrix, the closed form solution is still given by Equation~\eqref{eq:s_sol}. Note, however, that in this case the source matrix is level-dependent. For a given subdivision level $l$ and patch $p \in \{p_l\}$, $S^l_p$ is the concatenation of source matrices for all scales in $\{\mu_l\}$, as each shares the same source matrix. $S^l_p$ then has dimensions of ($N_\mathrm{S}$) by (number of pixels times number of scales in $\mu_l$), and the concatenated data matrix $X^l_p$ has dimensions (number of frequencies) by (number of pixels times number of scales in $\mu_l$).

For a column of the mixing matrix, the closed form now includes priors from the mixing matrix hierarchy:
\begin{align}\label{eq:a-sol}
a^i &= \left [\frac{R_is_i^T + \lambda_A \tilde{a}^i + \lambda_A \sum_j b^i_j}{||R_is_i^T + \lambda_A \tilde{a}^i + \lambda_A \sum_j b^i_j||_{F}} \right ]_+.
\end{align}
Here $\tilde{A}$ denotes the mixing matrix that defines the distribution from which $A$ is drawn. As an example, for $A = A^1_{p_1}$ this would be $\tilde{A}=A^0$. $B_j$ denotes the $j^\text{th}$ matrix that is drawn from $A$. As an example, for $A = A^1_{p_1}$ this would be $B_j=A^2_{p_1,j}$ for $j \in [1,4]$. The terms $b^i_j$ and $\tilde{a}^i$ are the $i^{\text{th}}$ column of $B_j$ and $\tilde{A}$ respectively. Finally, recall that $[\ldots]_+$ indicates a non-negativity constraint enforced by setting negative values to 0. Note that setting $\lambda_A$ to $0$ returns the GMCA update equation \eqref{eq:Fast-gmca-a-sol}.

At each level, we optimize the sources and mixing matrices iteratively before proceeding to the next level, iterating over all the levels until convergence is reached.

\subsection{CMB Prior}
While (H)GMCA is blind in theory, in practice it improves performance to include theoretical knowledge such as the frequency dependence of the CMB. \cite{Bobin} suggest enforcing this on the column of the mixing matrix corresponding to the CMB ($a^{\rm CMB}$) throughout optimization, which is equivalent to setting an infinitely strong (delta function) prior on the column. Because the CMB spectrum is not measured to infinite precision, in our GMCA\textsuperscript{LASSO} implementation we choose to use a Gaussian prior. For HGMCA, this means that the $a^{\rm CMB}$ column of the mixing matrix is drawn from this Gaussian prior rather than the hierarchy, reflecting the fact that the CMB has no local variation to the frequency dependence. The update equation for the CMB column then becomes

\begin{align}\label{eq:a-cmb-sol}
a^{\rm CMB} &= \left [\frac{R_is_i^T + \lambda_{\rm CMB} \tilde{a}^{\rm CMB}}{||R_is_i^T + \lambda_{\rm CMB} \tilde{a}^{\rm CMB}||_{F}} \right ]_+,
\end{align}
where $\lambda_{\rm CMB}$ is the strength of the CMB prior, and $\tilde{a}^{\rm CMB}$ is the theoretical frequency dependence of the CMB.
\subsection{Simulated Data and Wavelet Transformation}\label{sec:sim_data}

In order to understand the benefits of a hierarchical modeling strategy, we compare HGMCA with a number of baselines on simulated CMB and foregrounds. The simulations are generated using the Python Sky Model (\texttt{PySM}) code \citep{PySM} at a resolution of $N_{\rm side} = 256$ and bandlimited to $\ell_\text{CMB}=620$. \texttt{PySM} offers different models to simulate the foregrounds. These are built from \textit{Planck} templates and make conservative assumptions about their spectral dependence. We use Model 1 for synchrotron, thermal dust, anomalous microwave emission (AME), and free-free emission. The CMB random seed parameter was set to 2 for the final results and 1111 for hyperparameter tuning. The exact details of this model, its parameters, and the underlying assumptions can be found in \cite{PySM}. Note that while the \texttt{PySM} code is capable of modeling both instrument noise and distortions from the beam, for simplicity we do not include these effects in our simulated data. The maps are generated at frequencies corresponding to those measured by \textit{Planck} \citep{Planck2013Products}. In this work we focus on intensity maps, and leave the polarization maps for future work. 

The individual CMB and foreground maps that are used to create the mock full sky map for the results presented in this work are shown in Figure \ref{fig:inputs}. The bottom right panel shows the full sky image at 90 GHz. The top and bottom panels of Figure \ref{fig:PySM} show the ratio between the 90 GHz and 23 GHz maps for synchrotron radiation (top) and dust emission (bottom), which showcase the local variation in the foreground signal. If there were no spatial dependence to the frequency scaling, the ratio would be a constant value everywhere; instead, the ratio changes considerably as we move across the sky. We draw special attention to the spatial dependence of the frequency scaling because HGMCA has been built to better model local variations.

All of the inputs to the algorithms we present were converted into the scale-discretized wavelet basis. Appendix \ref{app:wavelet} describes this transform and the accompanying code in detail. For the results in this work we used $\lambda=3$, a bandlimit of $\ell_\mu=621$ (having a bandlimit one larger than the bandlimit of the data gives effectively zero reconstruction error), and an analysis depth for the scaling function of $J_0 = 1$. 
\begin{table*}
    \centering
\begin{tabular}{|l|l|l|l|l|}
\hline
Algorithm       & RMSE {\scriptsize $[\mu K]$}& RMSE\textsubscript{UT78} {\scriptsize $[\mu K]$}     & \% Error CMB $C_\ell$'s, all $\ell$'s & \% Error CMB $C_{\ell}'s, \ell>200$ \\ \hline
GMCA\textsuperscript{LASSO}            & 10.4430          & 1.2004   & 0.3815    &  0.1439   \\ \hline
HGMCA - 1 & 8.2860 & 1.1235 & 0.2582    & 0.0314\\ \hline
HGMCA - 2 & 7.7935 & 0.9044 & 0.1663    & 0.0322\\ \hline
HGMCA - 3 & 6.9835 & 0.9467 & 0.2311  & 0.0289\\ \hline
WavILC            & 4.3412          & 1.8538   & 0.9416    &  0.0487 \\ \hline
\end{tabular}
    \caption{Comparison of RMSE ($\mu K$) and average percent error in the CMB angular power spectrum including and excluding the \textit{Planck} UT78 mask region for WavILC, GMCA\textsuperscript{LASSO}, and HGMCA at several levels. GMCA\textsuperscript{LASSO} and HGMCA maps are selected via the input CMB blind criterion described in Section~\ref{sec:ran_seeds}. HGMCA outperforms GMCA\textsuperscript{LASSO} across all metrics and WavILC across all metrics except unmasked RMSE. The comparison between HGMCA-2 and HGMCA-3 is less clear, with HGMCA-2 winning on masked RMSE and overall $C_\ell$ error, and HGMCA-3 performing better on unmasked RMSE and larger $C_\ell$'s.}
    \label{table:results}
\end{table*}
\subsection{Wavelet ILC}\label{sec:imp_silc}
To compare HGMCA to the foreground correction algorithms used by \textit{Planck}, we have implemented a wavelet ILC (WavILC) code similar to what is done in \cite{planck2018foregrounds,NILC3,SILC}. We use the same scale-discretised wavelets as for our HGMCA and GMCA implementations, but the minimum $J_0$ is set to 6 (we find a larger analysis depth for the scaling function is important for ILC-type algorithms). To calculate the covariance matrix we follow the same approach as used in \cite{NILC3,SILC} and convolve our map with a Gaussian whose FWHM is set to be $9000/N_{\rm side}$ radians. More precise details of how the covariance matrix and weights are calculated can be found in Appendix \ref{app:NILC}. Note that both WavILC and the Needlet Internal Linear Combination (NILC) method on which it is based calculate different ILC weights for each pixel, meaning that they are capable of capturing local variation in the mixing matrix.\\

While our WavILC code uses the same scale-discretized wavelets as \cite{SILC}, our implementation has notable differences. \cite{SILC}'s implementation of Scale-discretized directional wavelet ILC (SILC) is unique among ILC algorithms in that it does not use the HEALPix pixelization scheme. Instead it uses the sampling scheme developed in \cite{ssht}, which requires fewer samples for a bandlimited signal. SILC also varies the wavelet dilation parameter $\lambda$ by multipole, whereas here we keep the same value of $\lambda=3$ for all multipoles. This is true both for the wavelet coefficients we use for WavILC as well as the wavelet coefficients we use for GMCA and HGMCA. Similarly, SILC takes advantage of the directional extension to scale-discrtized wavelets, while we do not. We also do not mask point sources or utilize inpainting because the simulations we present here are not as complex as the Full Focal Plane 8 simulations used by SILC \citep{FFP8}. 

\subsection{Pipeline}
The HGMCA results presented in this work were constructed through the following steps: 
\begin{enumerate}
    \item The simulation data is generated using \texttt{PySM}. The parameters for the PySM code are discussed in Section \ref{sec:sim_data}. The data is bandlimited to $\ell_\text{CMB}=620$.
    \item The region of the map corresponding to the \textit{Planck} UT78 mask is scaled by $0.5$ for all frequencies. We found that this preprocessing step improves convergence.
    \item The input data is transformed into the wavelet basis using the \href{http://www.s2let.org}{\texttt{S2LET}}\footnote{http://www.s2let.org} \citep{s2let1,mcewen2015ridgelet,mcewen2015novel,chan2016second} code. A discussion of the basis can be found in Appendix \ref{app:wavelet}. For the results in this work we used $\lambda=3$, a bandlimit of $\ell_\mu=621$, and a minimum $J_0 = 1$.
    \item HGMCA is run on the transformed input data until convergence. This consists of: 
    \begin{enumerate}
        \item Grouping the scales by their resolution limit. Since we do not have varying beam resolutions, this is dictated by the wavelet scales themselves (Section \ref{sec:HGMCA}), i.e. group data into the sets $\{\mu_l\}$.
        \item Initialize the mixing matrix hierarchy and the source matrix with one epoch of the base GMCA\textsuperscript{LASSO} algorithm. By providing a smart initialization, this step cuts back on the total runtime of HGMCA.
        \item For each epoch and each level, update the source and mixing matrices for each patch using Equations \eqref{eq:s_sol} and \eqref{eq:a-sol}. For results presented in Section \ref{sec:results}, we use the hyperparameter values $\lambda_S = 50$, $N_\mathrm{S}=5$, $\lambda_\text{CMB}$ = $10^{11}$, and $\lambda_A=5 \times 10^{10}$. A discussion of how these values are chosen and the sensitivity of our results to the hyperparameters can be found in Section \ref{sec:generalization}.
        \item Every 100 epochs, rather than using Equation \eqref{eq:s_sol} to update the sources, use the least squares solution $S=A_\text{pinv} X$ where $A_\text{pinv}$ is the pseudo-inverse of $A$. As we discuss in Section \ref{sec:imp_gmca}, we found empirically that this step improves convergence and helps avoid local minima.
    \end{enumerate}
    \item Use the mixing matrix hierarchy and the source matrix to reconstruct the CMB signal in the wavelet basis. If the known frequency dependence of the CMB was used as a prior, the column on which this prior was applied is selected as the CMB. Otherwise, whichever column most closely resembles the CMB frequency dependence is selected.
    \item Transform the wavelet representation of the CMB back into the pixel basis using \texttt{S2LET}.
\end{enumerate}

\section{Results}\label{sec:results}
The principal advantage of HGMCA is that it has the ability to fit to local variations without sacrificing the robustness offered by having access to the full map. To showcase the strength of the algorithm, we empirically compare its performance to both GMCA\textsuperscript{LASSO} (which produces one global fit) and WavILC (which can capture local variation). Using separate CMB sky realizations to tune parameters and evaluate results, we find that HGMCA is able to consistently return the highest fidelity CMB reconstruction (Section \ref{sec:alg_comp}). In addition, we show that GMCA\textsuperscript{LASSO} and HGMCA converge to a stable solution (Section \ref{sec:conv}), and confirm that our results hold under a number of generalization tests (Section \ref{sec:generalization}). Finally, we show that GMCA\textsuperscript{LASSO}, HGMCA, and WavILC do not introduce substantial non-Gaussianity into the reconstructed map, suggesting that such algorithms may be of interest for non-Gaussianity studies (Section \ref{sec:non-gauss}).

\subsection{GMCA\textsuperscript{LASSO}, HGMCA, and WavILC Comparison}\label{sec:alg_comp}   
We run a comparison between GMCA\textsuperscript{LASSO}, WavILC, and HGMCA at level 1 (HGMCA-1), level 2 (HGMCA-2), and level 3 (HGMCA-3)\footnote{The parameter values used for GMCA\textsuperscript{LASSO} and HGMCA are given in Appendix \ref{app:params} and the values for WavILC are discussed in Section \ref{sec:imp_silc}. The selection strategy for the parameters is discussed in Section \ref{sec:generalization} and Appendix \ref{app:params}.}. All of the algorithms are run on the input maps described in Section \ref{sec:sim_data}. We leave LGMCA out of the comparison because the code is not publicly available, and our implementation was not able to reproduce the improvements shown in \cite{L-GMCA}\footnote{The results presented in \cite{L-GMCA} are run on higher resolution simulations with additional contaminants and a varying beam.}. To determine the quality of the reconstructions, we use three metrics: (1) the root mean square error (RMSE) of the residual between the reconstructed output CMB and the input CMB maps, (2) the cross-power spectrum between the residual map and the input foregrounds, and (3) the difference between the power spectra of the input CMB and the reconstructed CMB.

\begin{figure*}
    \centering
    \includegraphics[scale=0.35]{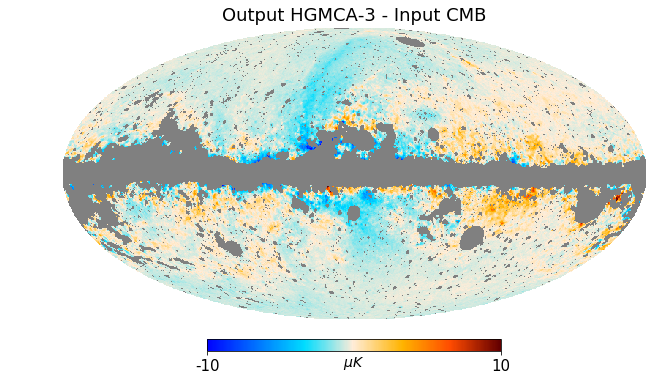}
    \includegraphics[scale=0.35]{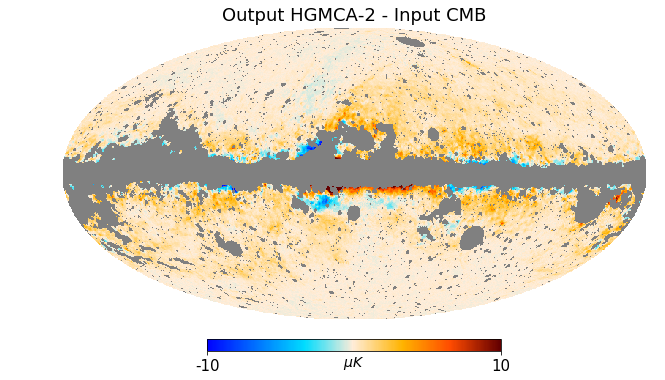}
    \includegraphics[scale=0.35]{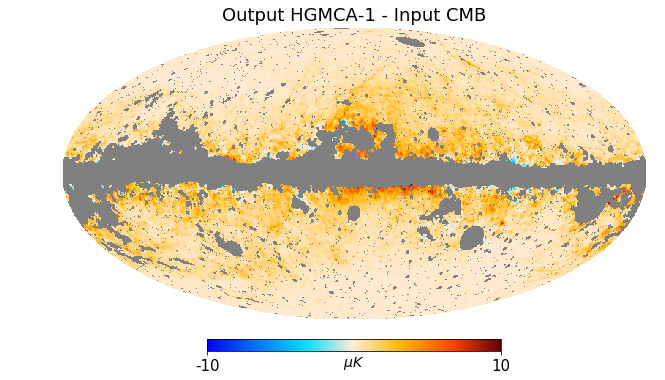}
    \includegraphics[scale=0.35]{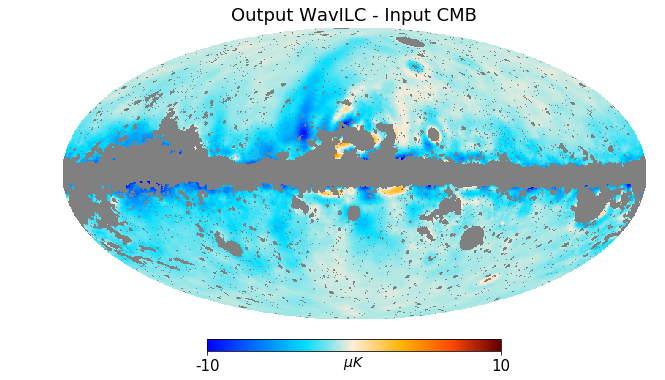}
    \includegraphics[scale=0.35]{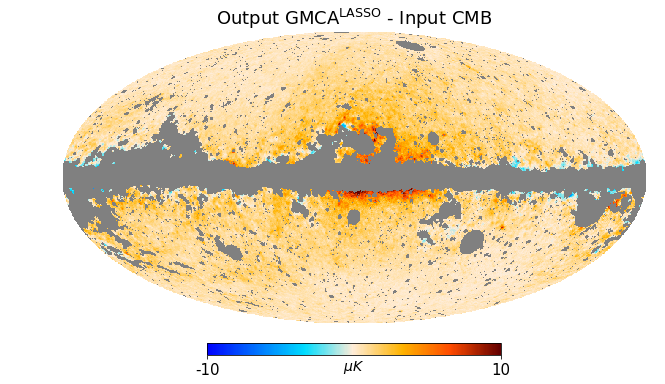}
    \caption{From top left to bottom: the residual maps for HGMCA-3, HGMCA-2, HGMCA-1, WavILC, and GMCA\textsuperscript{LASSO}. The dark gray region corresponds to the UT78 mask used by \textit{Planck}, and the scale of the color bar has been significantly reduced to accentuate the structure of the residual. All HGMCA maps have less residual foreground contamination than GMCA\textsuperscript{LASSO} and WavILC on the masked map, with HGMCA-2 as the lowest (see Table \ref{table:results}). }
    \label{fig:resHGMCA}
\end{figure*}
\begin{figure*}
    \centering
    \includegraphics[scale=0.35]{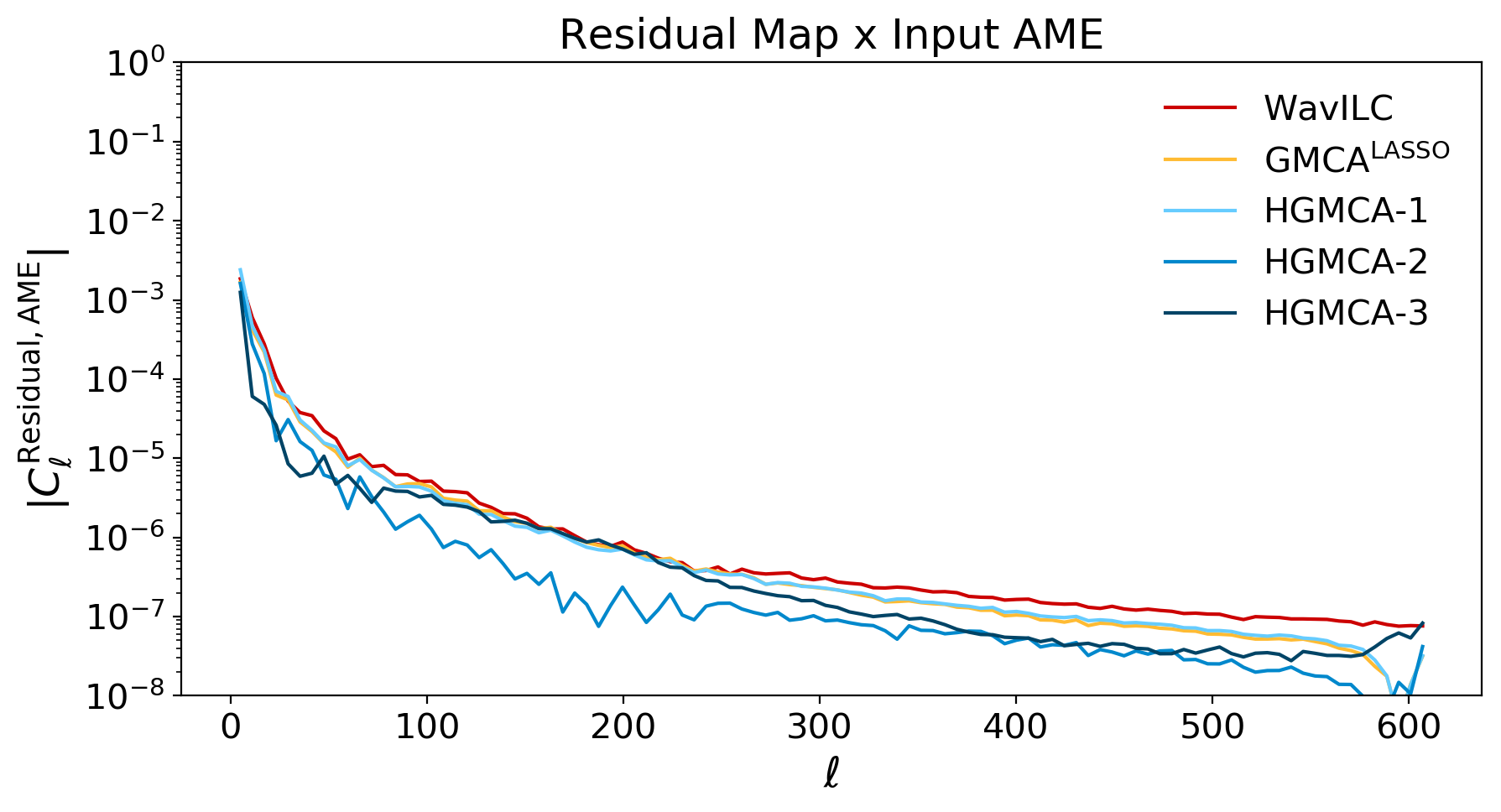}
    \includegraphics[scale=0.35]{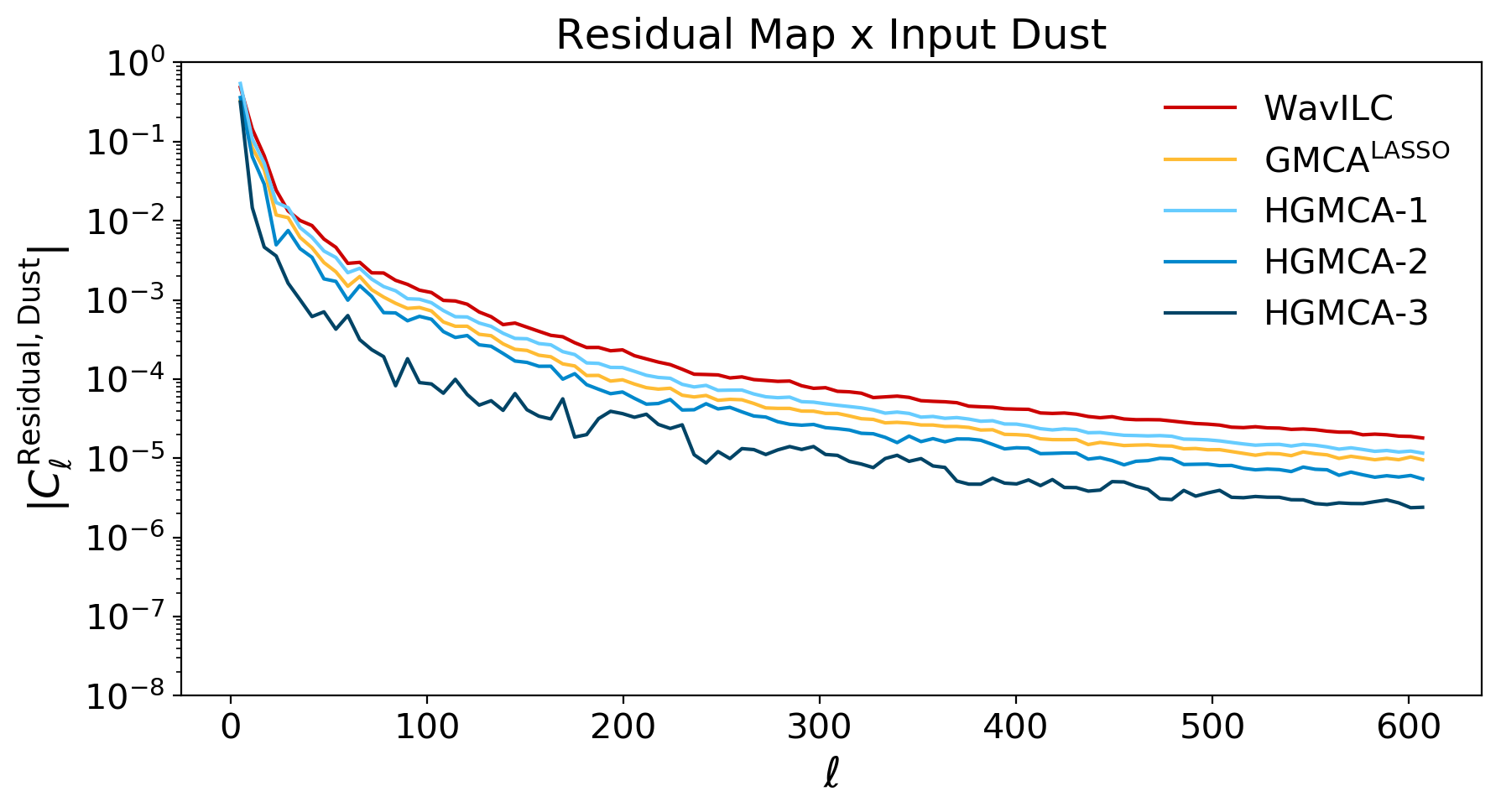}
    \includegraphics[scale=0.35]{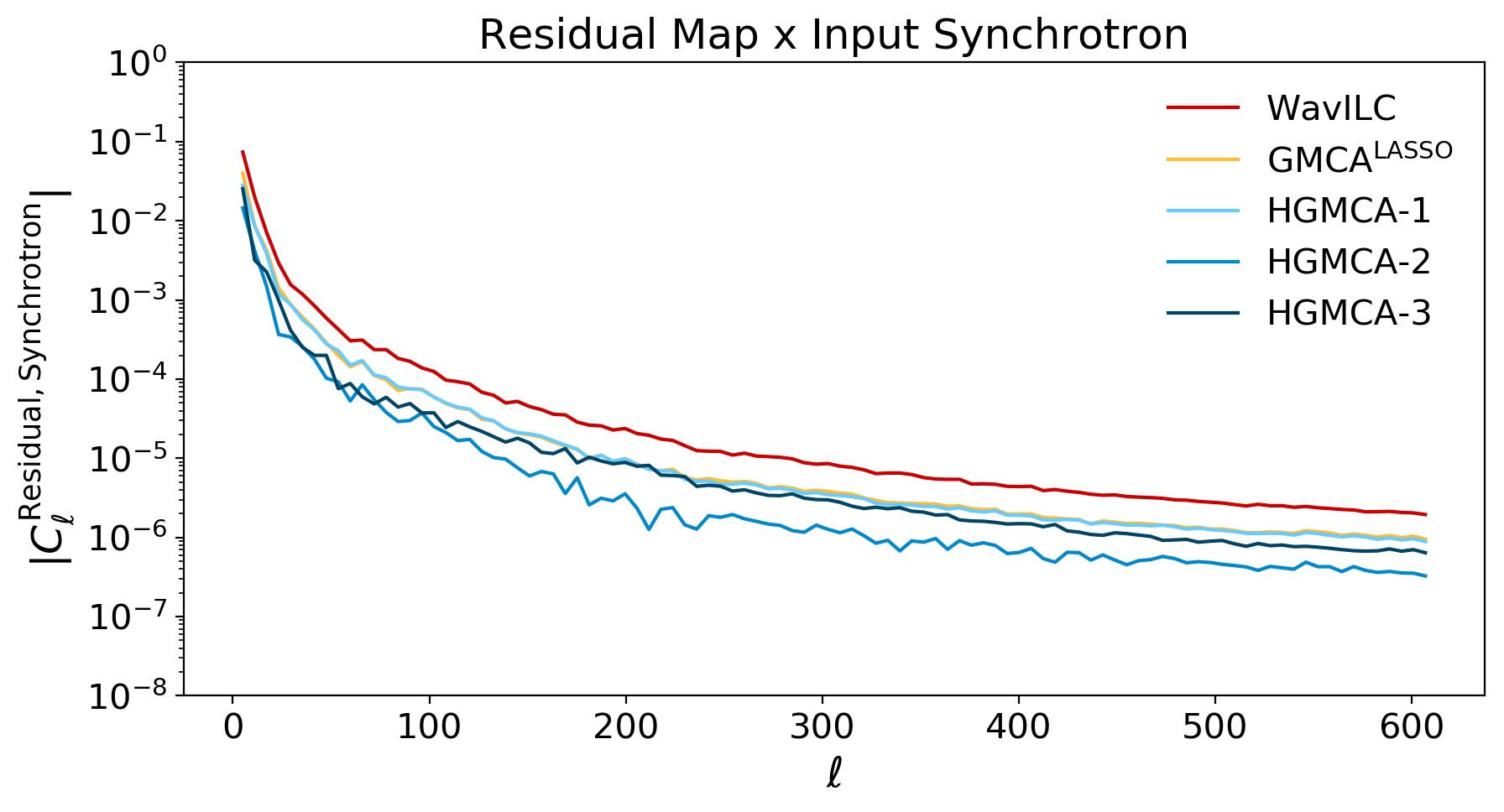}
    \includegraphics[scale=0.35]{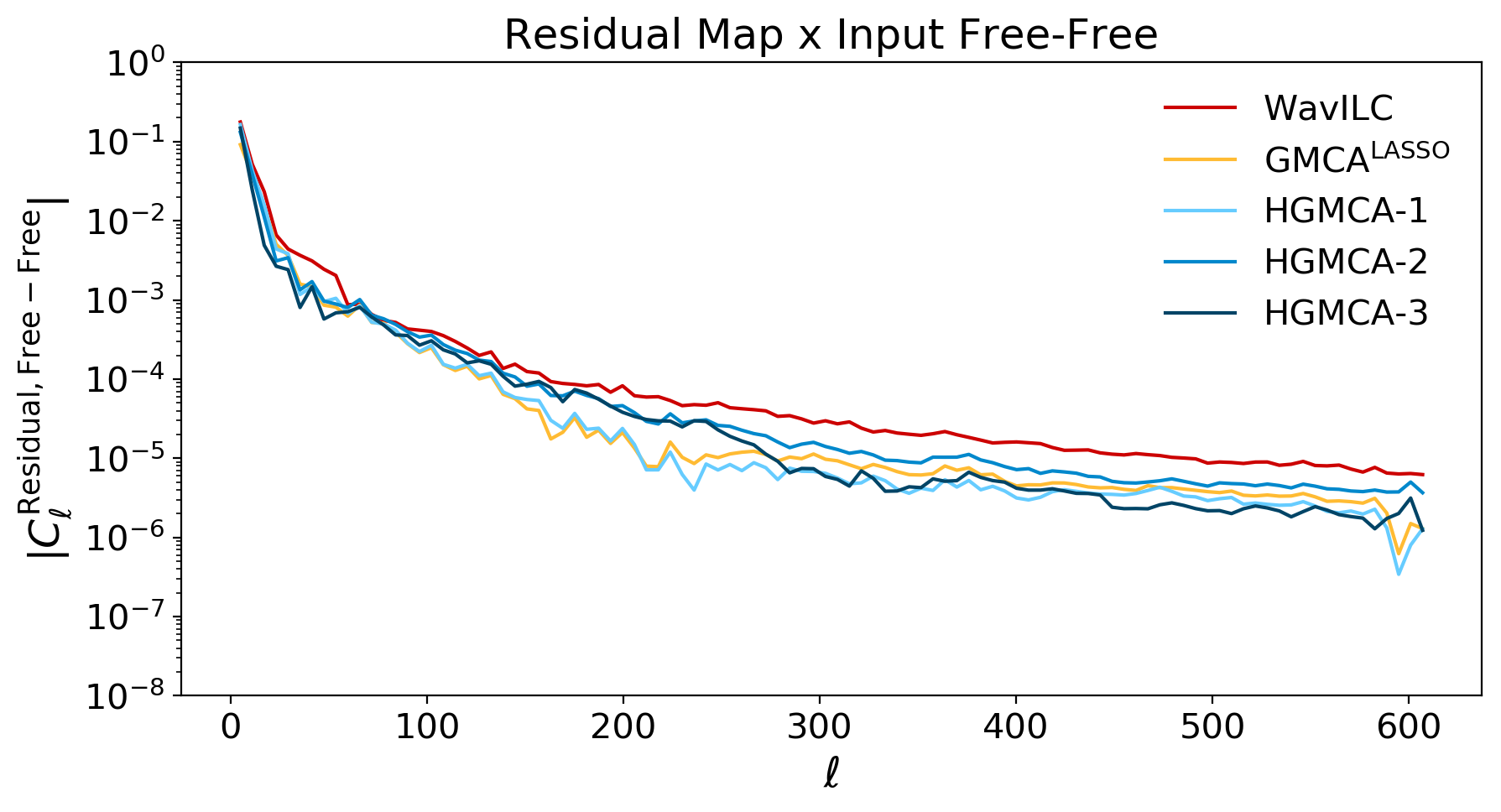}
    \caption{Magnitude of the cross-spectra between the residual CMB for each algorithm and the input AME, dust, synchrotron, and free-free foreground contaminants. A lower value of the cross-spectrum indicates a smaller level of contamination from that foreground component. HGMCA outperforms WavILC across all contaminants, especially at high $\ell$. HGMCA-1 shows similar performance to GMCA\textsuperscript{LASSO} throughout, and levels 2 and 3 outperform GMCA\textsuperscript{LASSO} across AME, dust, and synchrotron contaminants. In general, increasing the level of subdivision from $l=1 \to l=2$ minimizes foreground, but the step to $l=3$ does not. The only exceptions to this is the improved performance of HGMCA-3 on dust, and of GMCA\textsuperscript{LASSO} and HGMCA-1 on free-free contamination.}
    \label{fig:cl_contamination}
\end{figure*}

\begin{figure}
    \centering
    \includegraphics[scale=0.36]{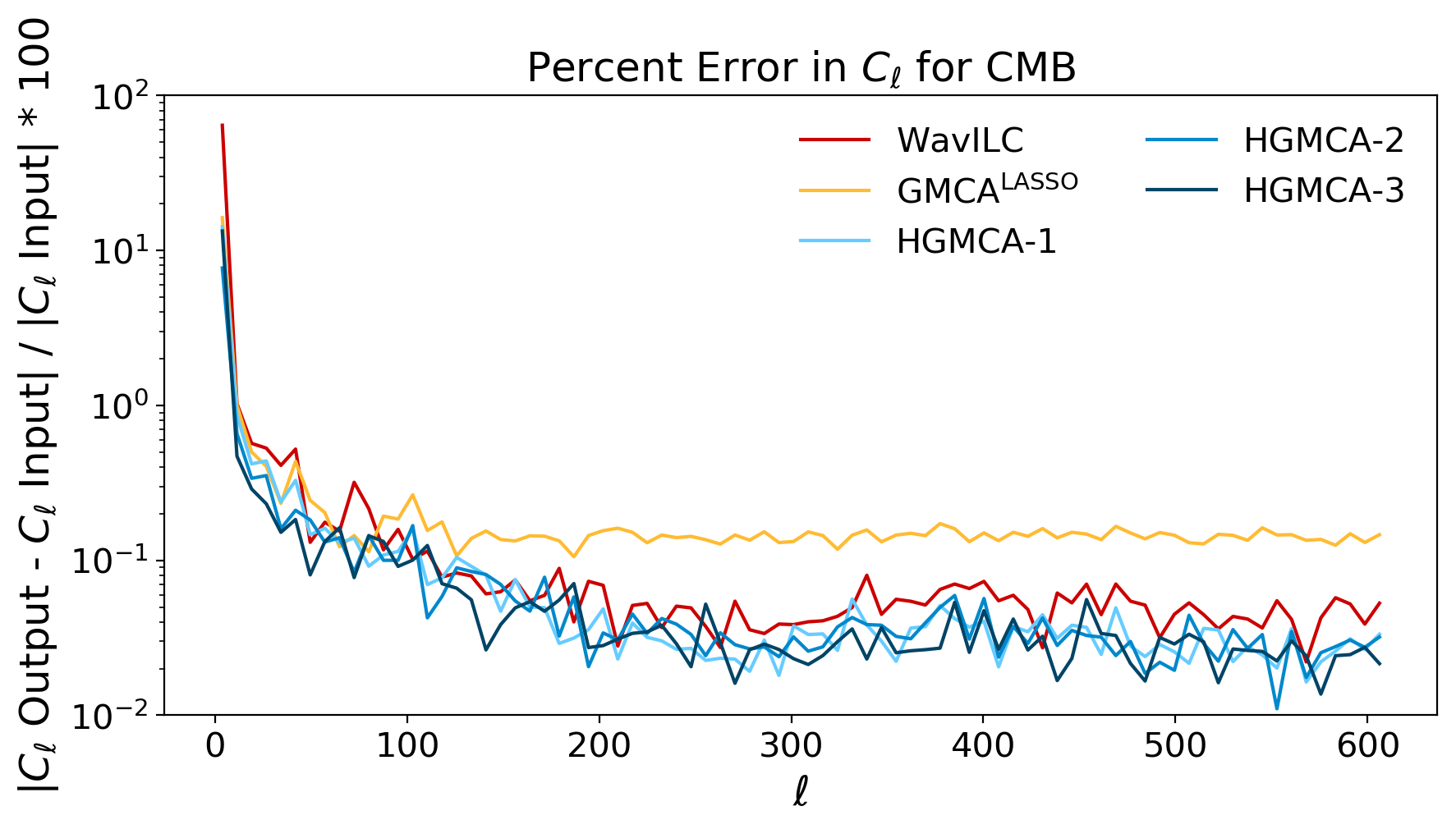}
    \caption{Percent error in $C_\ell$ as a function of multipole moment $\ell$ for each of the algorithms. WavILC shows its strength for reconstruction at high $\ell$ over GMCA\textsuperscript{LASSO}, but HGMCA is able to produce a better reconstruction across almost all values of $\ell$. The differences between the levels of HGMCA are more subtle, and are better understood quantitatively (see Table~\ref{table:results}).}
    \label{fig:cl_CMB}
\end{figure}

The results for each algorithm are presented in Table \ref{table:results}, where we show the RMSE between the input and output CMB maps (both with and without applying the UT78 mask used by \textit{Planck} (see \citealt{planck2015likelihoods})\footnote{Foregrounds typically dominate regions of the Galactic plane to such an extent that they cannot be removed when reconstructing the CMB from a few frequency channels (many more channels, as well as external data, would be needed to successfully disentangle the CMB from foreground in those regions). Therefore, most applications (\eg measurements of non-Gaussianity, isotropy, CMB lensing, \citealt{planck2013isw, planck2018lensing, planck2018isotropy,  planck2018png}) conservatively mask those regions.}, as well as the difference in their power spectra. The HGMCA algorithm generates a better reconstruction of the CMB than GMCA\textsuperscript{LASSO} across all metrics and WavILC across all metrics except unmasked RMSE. In Figure \ref{fig:resHGMCA} we present the masked residual maps for the five algorithms. The residuals show that all three levels of HGMCA offer a marked improvement over GMCA\textsuperscript{LASSO} and WavILC across the masked map. Notably, HGMCA shows little residual signal at the poles of the map, with almost none at all for HGMCA-2. For the HGMCA maps, the masked RMSE is roughly equivalent for levels $l=3$ and $l=2$, but the residual increases as we go to level $l=1$. Figures \ref{fig:resHGMCA} and \ref{fig:cl_contamination} show that this increased RMSE comes from additional dust contamination (see Figure \ref{fig:inputs} for the shape of the dust foreground).

To further understand the remaining foreground contamination, we show the correlation coefficients between the residuals of the five reconstructed maps (GMCA\textsuperscript{LASSO}, WavILC, HGMCA-1, HGMCA-2, and HGMCA-3) and the input foregrounds in Figure \ref{fig:cl_contamination}. By using the residual maps, we can probe what contaminants still remain in the reconstructed maps and at which multipoles they contribute. For anomalous microwave emission (AME), dust, and synchrotron contamination, HGMCA shows a reduction by an order of magnitude over GMCA\textsuperscript{LASSO} and WavILC. These differences in source and level of contamination are further evidenced by the residual maps in Figure~\ref{fig:resHGMCA}. Especially noticeable is the difference in the residual maps of WavILC and HGMCA-2. WavILC's residual map shows clear resemblance to the input synchrotron foreground from Figure \ref{fig:inputs}, whereas HGMCA-2's residual contamination is focused near the core, likely coming from dust or free-free emissions. Similar structure in the residuals was also observed in the application of other ILC implementations to the \textit{Planck} simulations \citep{SILC}. Large-scale residuals in the CMB reconstructions are typical of ILC-type methods since they result from chance correlations between the CMB and the foregrounds and the variance minimization built into the algorithm. This unintentionally cancels a few modes, proportionally to the power spectrum of the CMB, and the effect is therefore more pronounced on large scales. The improvement of HGMCA over GMCA\textsuperscript{LASSO} seems to come mostly from decreasing the contamination from AME and synchrotron at the second level, and dust at the third level.

Within the different levels of HGMCA itself, as the error in the $C_{\ell}$'s in Table~\ref{table:results} suggest, HGMCA-2 has the lowest correlation to AME and synchrotron contaminants, and the lowest overall contamination. Our analysis also suggests that the improvement of HGMCA-3 over HGMCA-1 comes from an order of magnitude improvement on dust contamination (reflected in the residual map itself in Figure~\ref{fig:resHGMCA}) and the improvement at low multipoles. Despite this large improvement in dust contamination, HGMCA-3 has increased AME and synchrotron residuals when compared to HGMCA-2, causing HGMCA-2 to win across most metrics.

We also look at the power spectrum difference between the input CMB and reconstructed CMB for all five algorithms. Figure \ref{fig:cl_CMB} shows the percent level error as a function of $\ell$. As with the previous two tests, HGMCA at all levels offers a significant improvement over the base GMCA\textsuperscript{LASSO} algorithm. This difference is most pronounced at $\ell>300$, but persists all the way to the smallest $\ell$. WavILC is generally competitive with HGMCA, except in the lowest $\ell$ bin, where it is an order of magnitude larger (this causes the average $C_{\ell}$ error to be higher). Of the three HGMCA variants, HGMCA-2 performs the best overall, with a $\sim0.065\%$ improvement over the second best performer, HGMCA-3. At mid to high $\ell$'s, HGMCA-3 slightly improves over HGMCA-2 by about $0.003\%$. GMCA\textsuperscript{LASSO} performs poorly on the $C_\ell$'s because it struggles to converge to the correct mixing matrix column for the CMB. Post-processing the GMCA\textsuperscript{LASSO} solution by enforcing that this column equal the CMB prior makes GMCA more competitive on the $C_\ell$'s but has a negligible impact on other metrics.

All three metrics show that the added versatility permitted by the HGMCA model leads to lower foreground contamination in CMB reconstructions. Interestingly, while the transition from GMCA to HGMCA-1 to HGMCA-2 leads to a consistent improvement in results, HGMCA-3 does not improve over HGMCA-2. This is likely because the additional local variation captured in transition between level 2 and level 3 of subdivision is small, and the low resolution of our maps limits how deep we can go before the optimization at the deepest levels is prior dominated. The generalization tests in Section \ref{sec:for_temp} show how increasing the local variation at smaller scales causes HGMCA-3 to outperform HGMCA-2.

\subsection{Convergence}\label{sec:conv}
\begin{figure}
    \centering
    \includegraphics[scale=0.4]{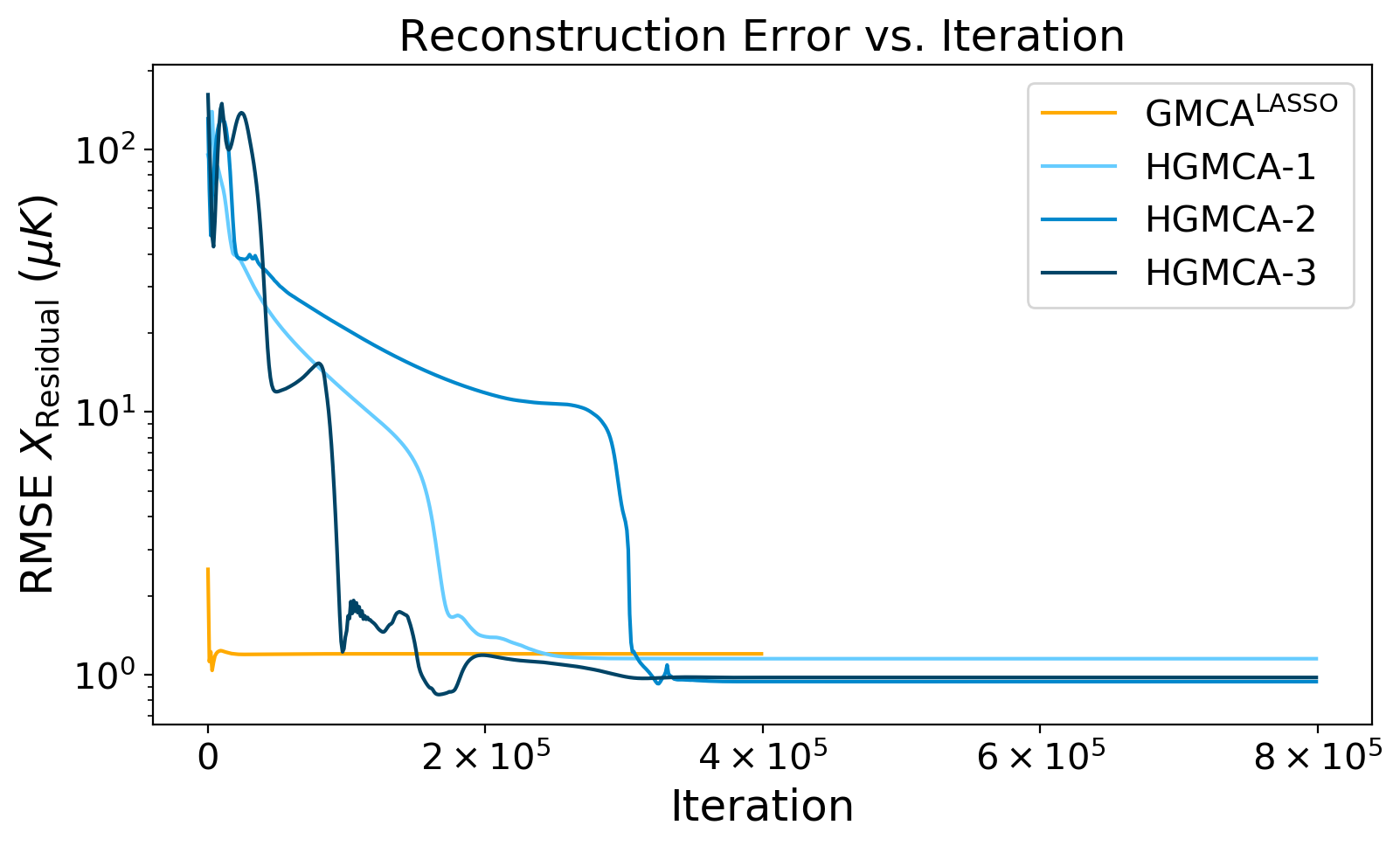}
    \caption{The reconstruction error as a function of iteration for the HGMCA and GMCA\textsuperscript{LASSO} maps presented in Table \ref{table:results}. Because we are measuring the RMSE of the residual map, we do not expect the final converged RMSE to be the lowest. However, all four algorithms converge in the limit of large numbers of iterations.}
    \label{fig:conv}
\end{figure}

In Figure \ref{fig:conv} we show that the reconstruction error of GMCA\textsuperscript{LASSO} and HGMCA converge in the limit of many iterations. Here, reconstruction error is measured as the root mean squared error (RMSE) of the residual map. Note that we are measuring the convergence of the error on the residual map (which we cannot directly measure in our loss function), so we do not expect it to monotonically decrease. Indeed, at some points our algorithm achieves lower RMSE than its converged value. However, in order to make our algorithm blind to the input CMB, we always report on the converged map. This ensures that our results may still be applied in practice, where the input CMB is unknown.

Since our implementations of both algorithms reach a stable state after sufficient iterations, we do not need to tune any specific convergence parameter to achieve good performance. In this paper, all of the HGMCA results have been run for 800,000 iterations and all of the GMCA\textsuperscript{LASSO} results have been run for 400,000 iterations. These values were chosen to be well beyond the point of convergence for both algorithms and therefore ensure that all of our results are stable.

\subsection{Generalization Errors}\label{sec:generalization}

\begin{figure*}
    \centering
    \includegraphics[scale=0.35]{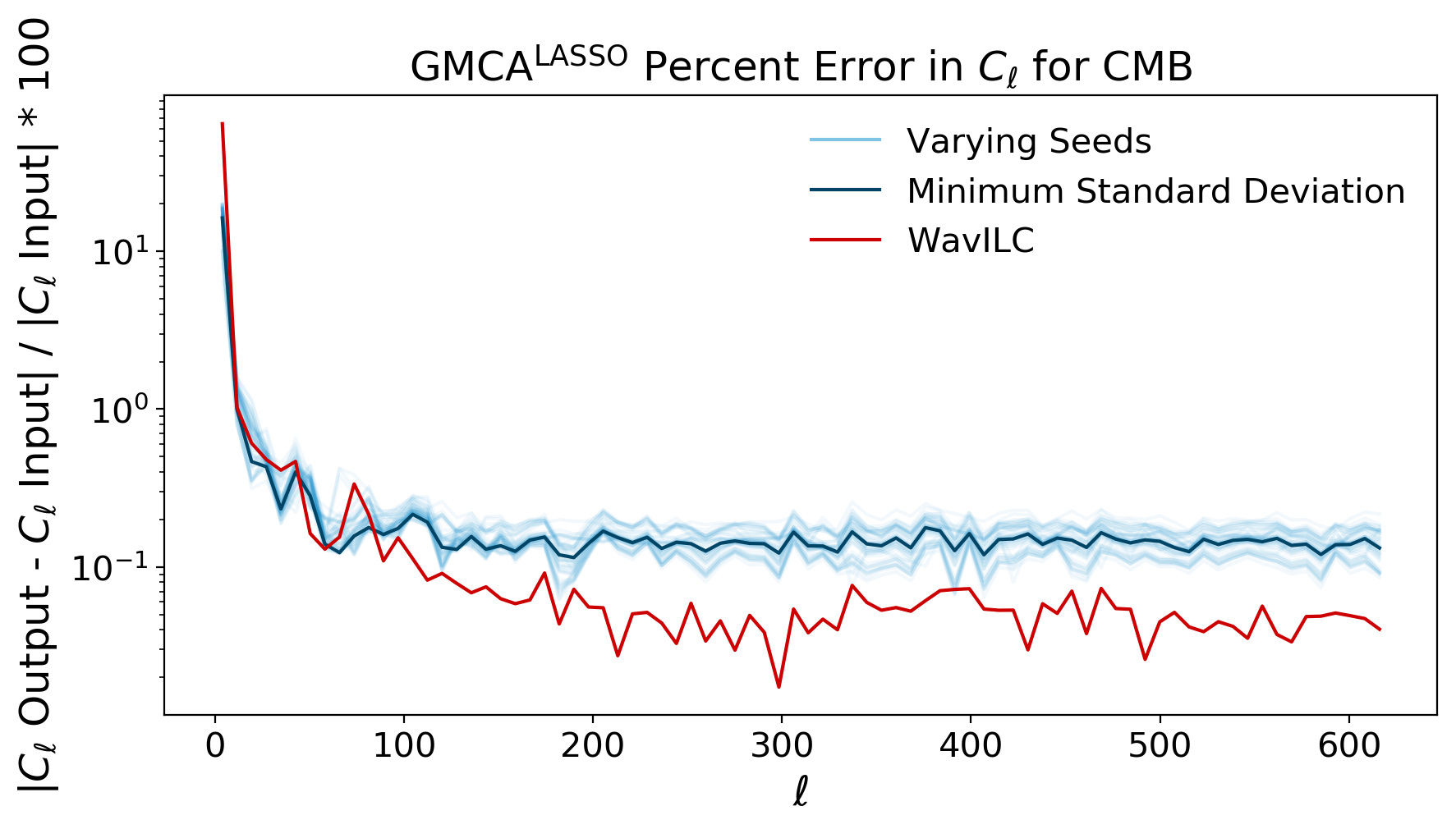}
    \includegraphics[scale=0.35]{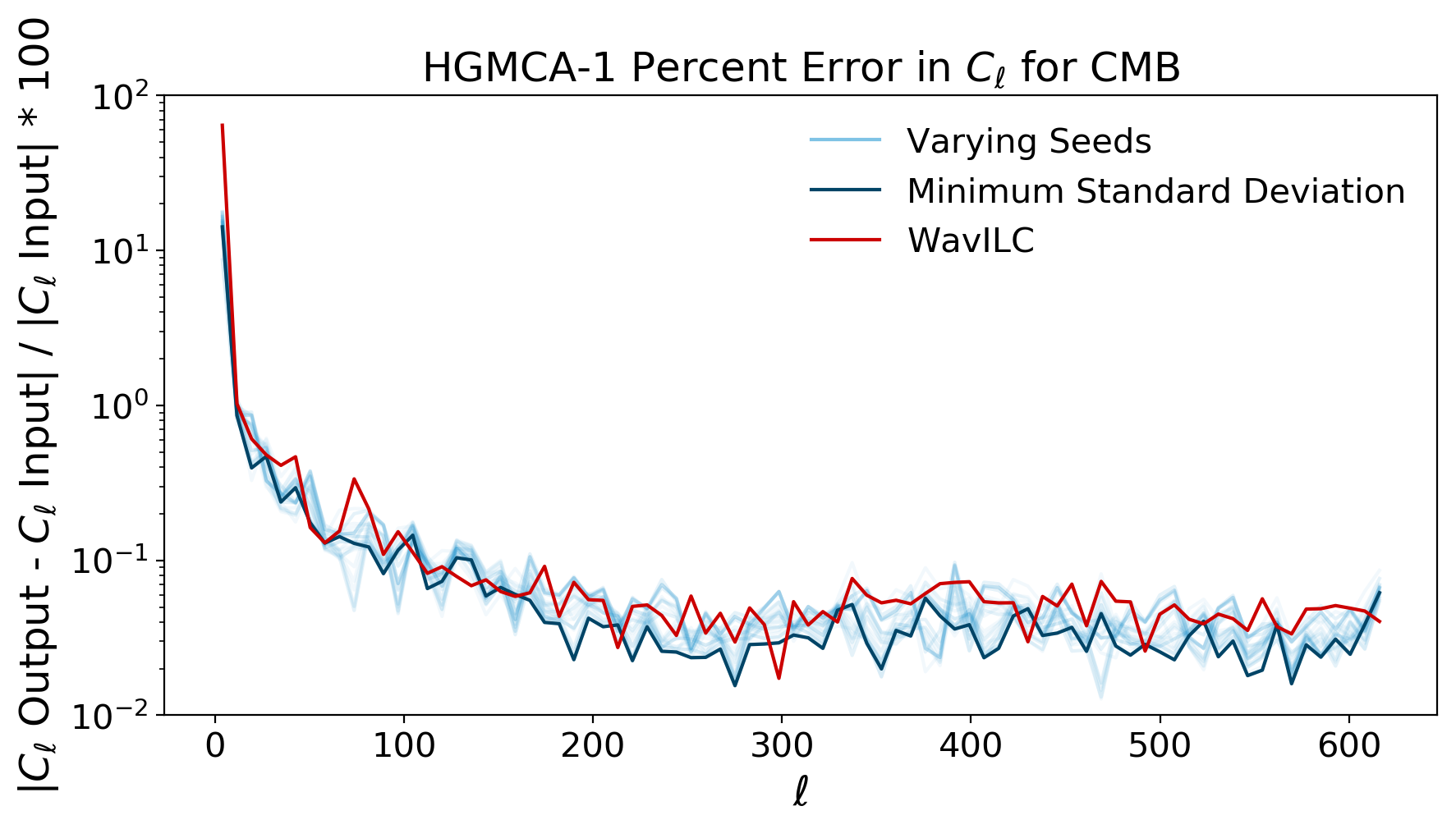}
    \includegraphics[scale=0.35]{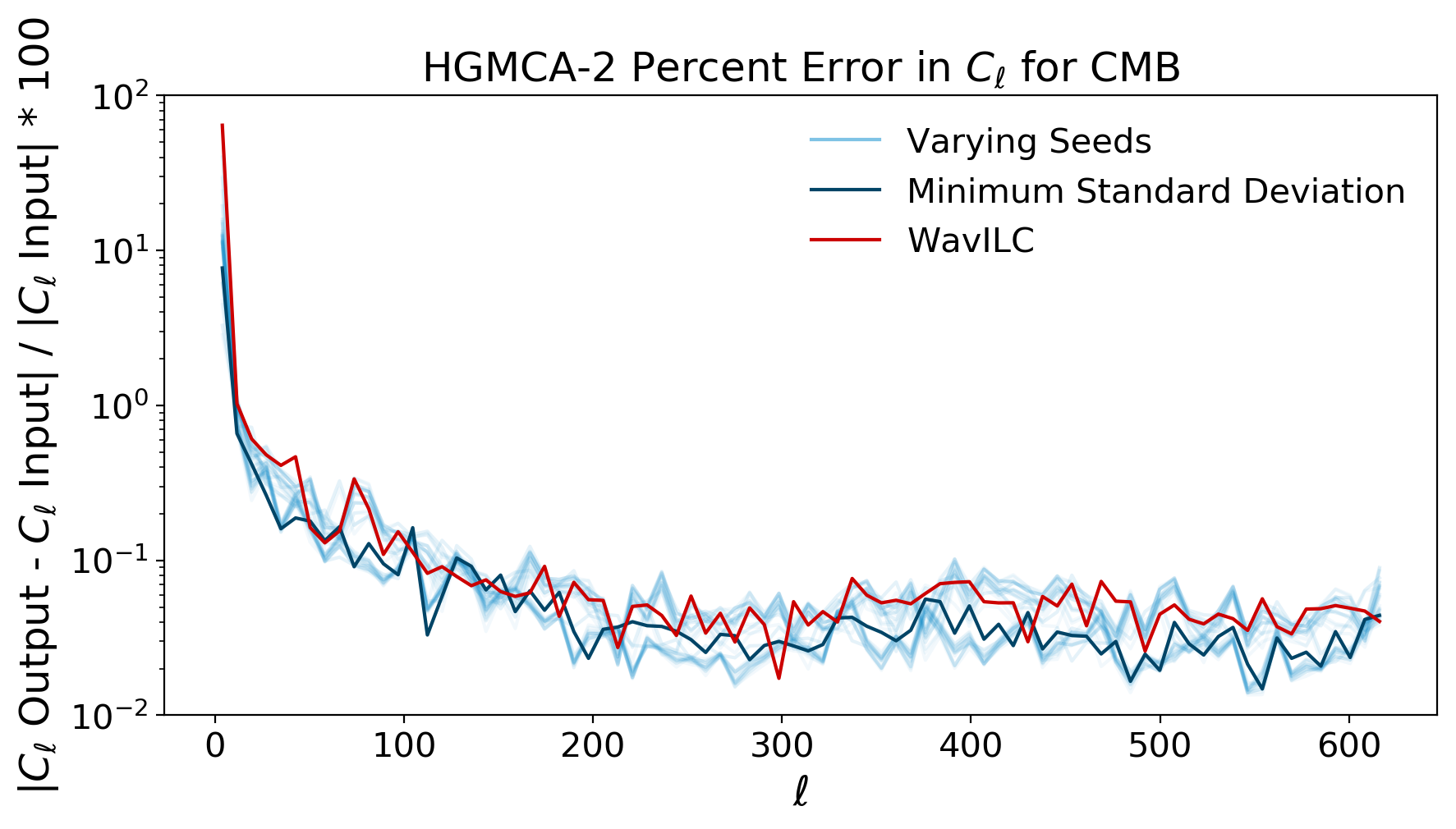}
    \includegraphics[scale=0.35]{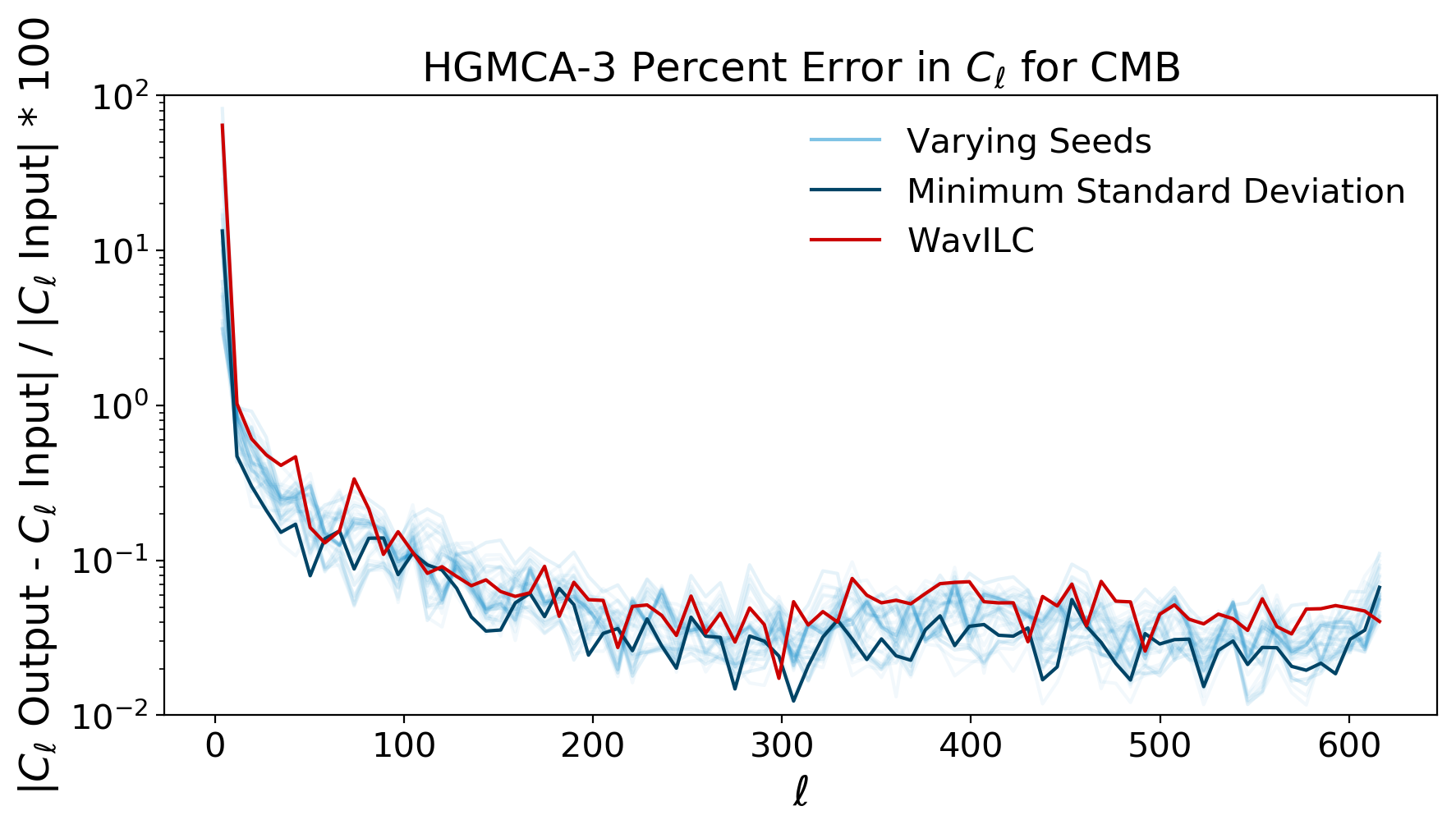}
    \caption{Percent error in the CMB $C_\ell$'s as a function of $\ell$ for varying seeds of each of the algorithms. The transparent blue lines are different random seeds with the same parameter configuration, the red lines are the results for WavILC, and the dark blue lines are the results we present in Section \ref{sec:results}. The selection function used to pick the maps that return the dark blue lines is blind to the input CMB; it is explained in detail in section \ref{sec:ran_seeds}. Note that the dark blue lines (which correspond to our fiducial results) for all four algorithms is not an outlier compared to other random seeds.}
    \label{fig:generalization}
\end{figure*}
\begin{figure*}
    \centering
    \includegraphics[scale=0.35]{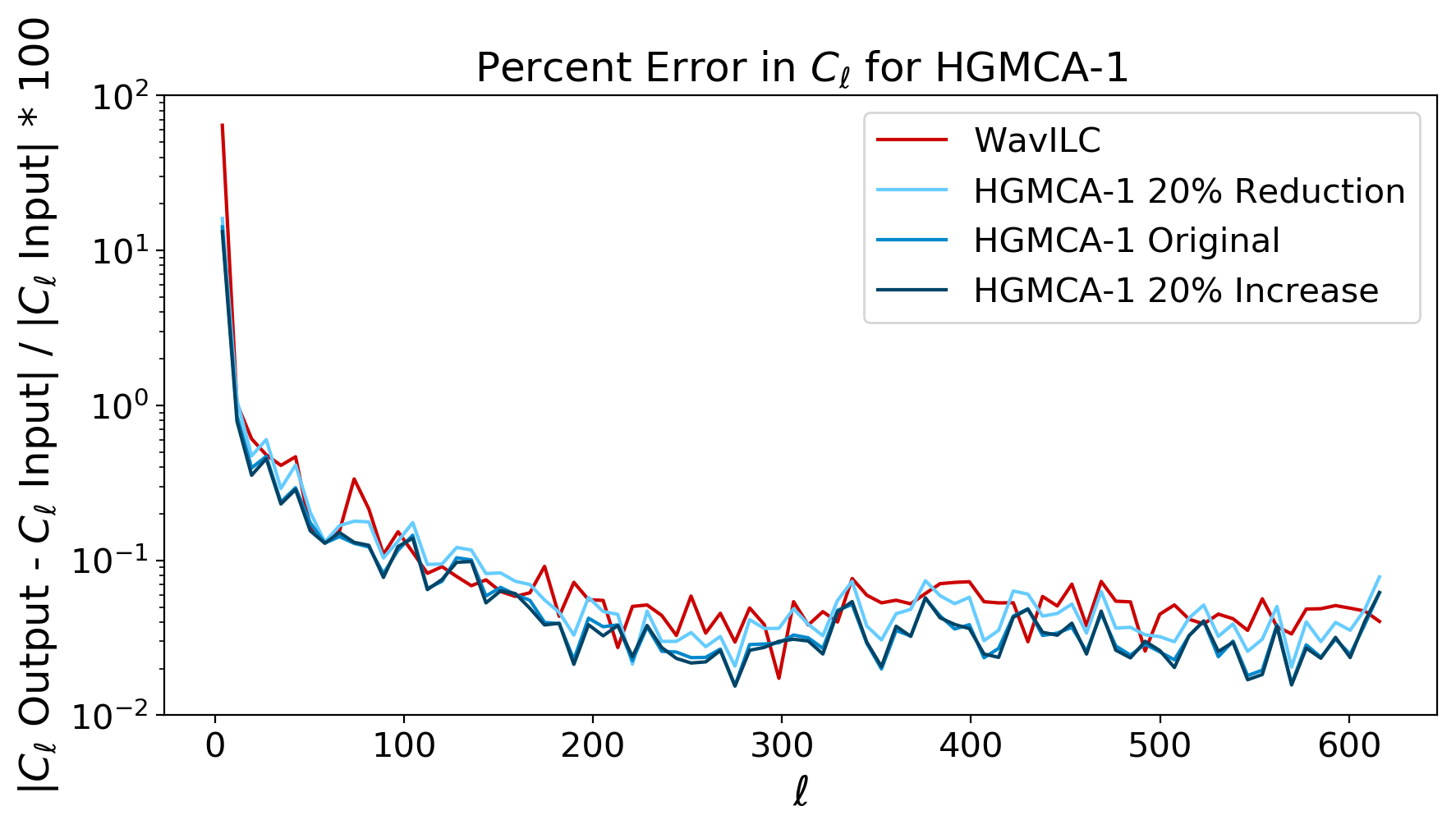}
    \includegraphics[scale=0.35]{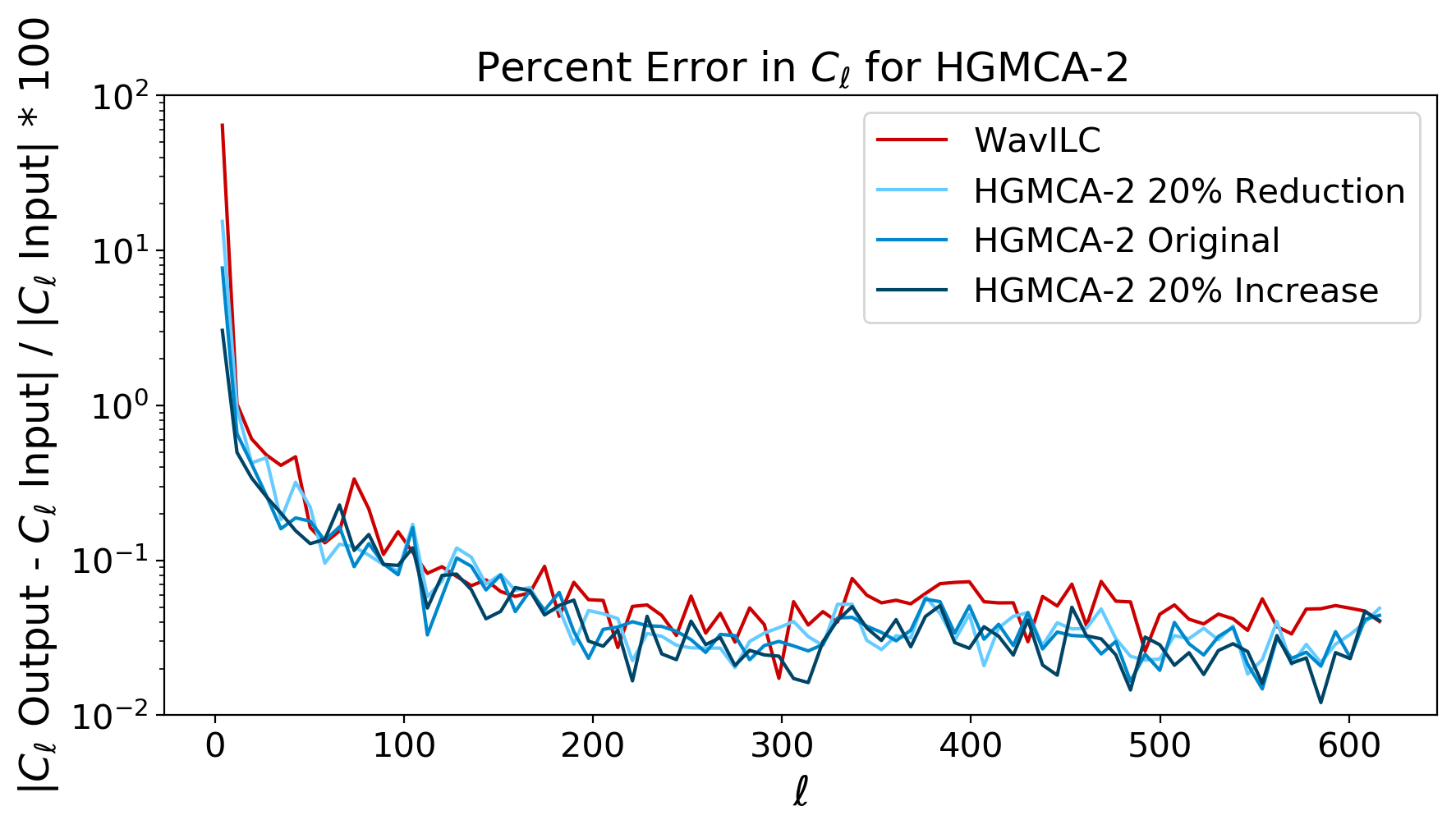}
    \includegraphics[scale=0.35]{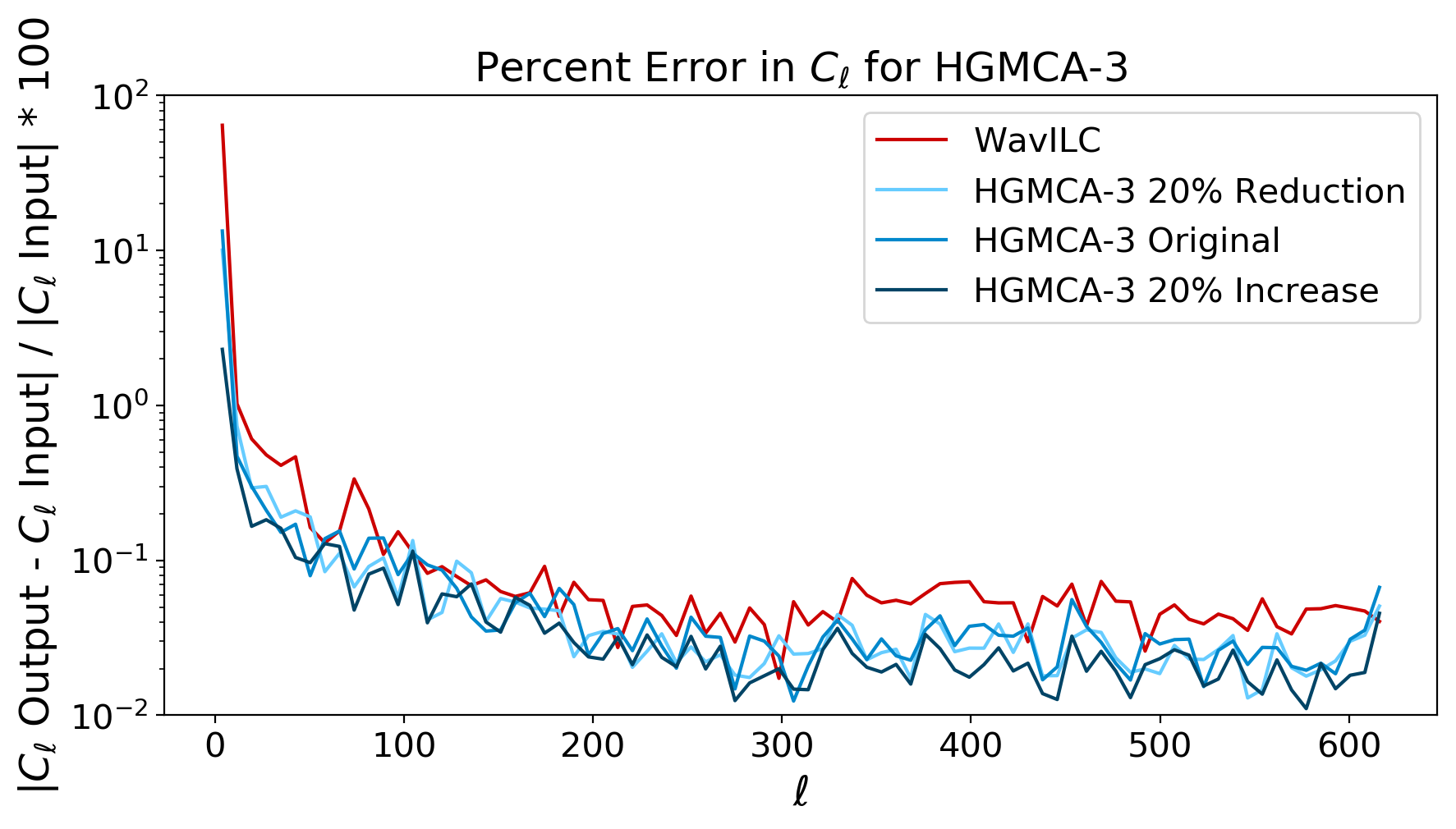}
    \caption{Percent error in the CMB $C_\ell$'s as a function of $\ell$ for varying parameter configurations. For all three HGMCA variants we increase and decrease all parameters by 20\% to see how it affects the performance of the algorithm. As with the random seeds, the results we present are in line with what other parameter configurations return. }
    \label{fig:param_var}
\end{figure*}
\begin{figure}
    \centering
    \includegraphics[scale=0.35]{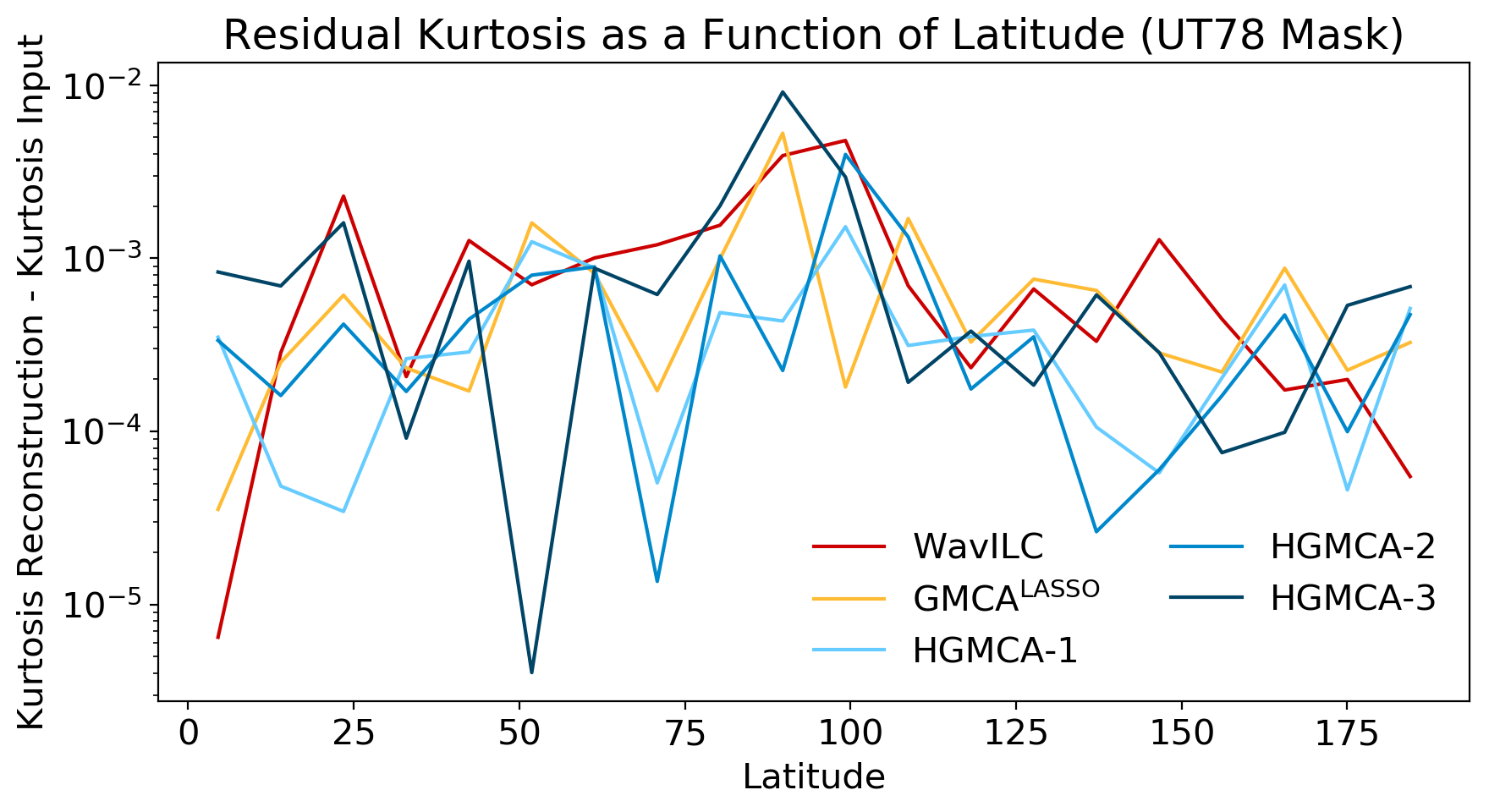}
    \caption{The difference between the kurtosis of the input CMB and the kurtosis of the outputs. None of the five algorithms has a clear advantage over the others, but all five do seem to introduce minimal kurtosis in their residuals. Note that the UT78 \textit{Planck} mask has been applied to the maps (meaning that the regions within the mask were not used for the kurtosis calculation).}
    \label{fig:non_gauss}
\end{figure}

There are four parameters intrinsic to the HGMCA algorithm that must be set either theoretically or empirically: $\lambda_S$, $\lambda_A$, $\lambda_\text{\rm CMB}$, and $N_\mathrm{S}$. We ran a grid search over a wide range of possible values for the first three of these parameters and selected the configuration that yielded the lowest RMSE error for GMCA\textsuperscript{LASSO} and HGMCA (the exact values can be found in Appendix \ref{app:params}). One potential concern is that the performance achieved by these parameter configurations is not generalizable beyond the simulations used to select them; this effect is often called overfitting. In this section, we discuss the four major steps we have taken to address this issue: 1) the independence of training and testing simulations, and the robustness of HGMCA 2) to random seeds, 3) to perturbations in parameters, and 4) to changes in foreground simulation.

\subsubsection{CMB Realization}

In order to show that our results are not an artifact of the specific simulation we used to tune our parameters, we exclusively report on a different simulation set. \texttt{PySM} generates the CMB from a Gaussian random field, and thus returns different maps depending on the seed for this process. For our hyperparameter tuning, this seed was set to 1111 instead of 2, as reported in Section \ref{sec:sim_data}. By changing the CMB instance, we train our parameters and report results on different maps, thus showing the robustness of our algorithm to changes in CMB simulation.

\subsubsection{Random Seeds}\label{sec:ran_seeds}
One concern with the coordinate descent optimization strategy we employ for GMCA\textsuperscript{LASSO} and HGMCA is that the order in which the sources and levels are updated is arbitrary. We do not want the results of our algorithm to be strongly dependent on the specific order, since that too would be a sign of poor generalization. The standard approach is to choose the order of the sources or levels randomly at each step of optimization. This allows us to probe a wide range of orders simply by altering the seed used by the pseudo-random number generator of our algorithm. In Figure \ref{fig:generalization} we show the results for GMCA\textsuperscript{LASSO}, HGMCA-1, HGMCA-2, and HGMCA-3 on the percent error in the CMB $C_\ell$ metric for a number of random seeds. For each of the four plots, the transparent blue lines correspond to the results of different seeds. We are only showing the results of runs whose standard deviation of the reconstructed map falls in the bottom 75\% of the random seeds to remove outliers. The red line corresponds to the performance of WavILC and the dark blue line corresponds to the maps we presented as our results in Section \ref{sec:results}. 

The maps we used for our results were selected by making two cuts. First, as we mentioned in Section \ref{sec:imp_gmca}, every 100 iterations of our algorithm we conduct a single step of least squares optimization on the source. This is equivalent to setting
\begin{align}
    S = A_\text{pinv} X,
\end{align}
where $A_\text{pinv}$ is the Moore-Penrose pseudo-inverse of the matrix $A$. Since the matrix $A_\text{pinv}$ is not a true inverse, $AS = A A_\text{pinv} X$ may not equal $X$. This means that some $A$ matrices will return a large error for $||X - A A_\text{pinv} X||^2_F$ even without sparsity constraints, suggesting they will return a lower quality reconstruction. We therefore conservatively cut any run whose pseudo-inverse introduces error in the top 50\% of runs. From the remaining runs, we select the one with the minimum standard deviation of the reconstructed map. Note that both of these cuts are completely blind to the input CMB map and are therefore reproducible outside of simulations. 

While employing these two cuts does help select one of the better performing seeds, Figure \ref{fig:generalization} shows that selecting any seed with small standard deviation (in the bottom 75\% of seeds) has better or equivalent performance to WavILC. This demonstrates that the results presented in this work are robust to the exact pattern of optimization that is employed. While the selection function we use for our results does extract one of the seeds with better performance, it is neither an outlier nor the best performer.

\subsubsection{Parameter Perturbations}
While we have made efforts to ensure that the parameter values used for our results are not overfit to the specific simulation, it is still important to understand the sensitivity of our simulations to parameter choices. In Figure \ref{fig:param_var} we show how our results are affected by a 20\% decrease and a 20\% increase in parameter values for all three levels of HGMCA. The lines plotted here were extracted using the same selection function described in Section \ref{sec:ran_seeds} from a number of random seeds. As with the random seed test, there does not appear to be a significant change in performance even when our parameters are all simultaneously changed by a significant percentage. The results we present are also not an outlier; in fact, increasing all the parameter values by 20\% seems to slightly improve our performance on this particular simulation.

\subsubsection{Foreground Templates}\label{sec:for_temp}
We have shown that our results are robust to different instantiations of the CMB as well as perturbations in parameters and optimization, but we have not yet shown how HGMCA performs under different types of local variation and levels of sparsity in the foreground contaminants. To show this, we employ \texttt{PySM}'s Model 2 templates for dust and synchrotron emissions, which give more sophisticated models of local variation for these two sources. Especially noticeable is the fact that the dust variation is of a smaller scale, and cannot be captured until deeper levels of subdivision. We therefore expect not to see a noticeable improvement between GMCA\textsuperscript{LASSO} and lower-level HGMCA. In addition, Model 2 changes the L1-norm (relaxed sparsity) of these sources in the wavelet space by $15-20\%$. To show the robustness of HGMCA to these changes, we continue to use the parameters trained on Model 1 blind to the new templates. However, because of the smaller scale variation in dust, we also present results on HGMCA at level 4 (HGMCA-4).

Table~\ref{table:model-2} shows a comparison between the performance of HGMCA, GMCA\textsuperscript{LASSO}, and WavILC on these simulations with respect to RMSE and difference in power spectra between the input and output maps. Table~\ref{table:cor-model-2} compares the correlation of each algorithm's residual error (the difference of input and output CMB) to each foreground source. Detailed figures of these comparisons can be found in Appendix~\ref{app:gen}. All error metrics show that HGMCA-1 and HGMCA-2 struggle with the smaller scale local variation of model-2, and underperform when compared even to GMCA\textsuperscript{LASSO}. While HGMCA-1 and HGMCA-2 must have optima that match GMCA\textsuperscript{LASSO}, with little local variation to take advantage of at levels 1 and 2, it is expected that the simpler optimization space of GMCA\textsuperscript{LASSO} will provide a better solution in practice. However, HGMCA-3 and HGMCA-4 are at a depth at which it is possible to take advantage of local variation and show a marked improvement across almost all error metrics over HGMCA-1, HGMCA-2, and GMCA\textsuperscript{LASSO}. Overall, HGMCA-4 and WavILC perform comparably, with WavILC providing better RMSE and $C_\ell$'s for $\ell>200$, and HGMCA-4 providing lower residual contamination and $C_\ell$ error for low $\ell$.

\begin{table*}
    \centering
\begin{tabular}{|l|l|l|l|l|}
\hline
Algorithm       & RMSE {\scriptsize $[\mu K]$} & RMSE\textsubscript{UT78} {\scriptsize $[\mu K]$}     & \% Error CMB $C_\ell$'s, all $\ell$'s & \% Error CMB $C_{\ell}'s, \ell>200$ \\ \hline
GMCA\textsuperscript{LASSO}            & 23.4081          & 4.5756   & 81.7473    &  0.5335 \\ \hline
WavILC            & 6.1665          & 1.4835   & 8.7473    &  0.04399\\ \hline
HGMCA - 1 & 35.0097 & 8.9987 & 382.5847    &  0.2703\\ \hline
HGMCA - 2 & 33.5835 & 8.8376 & 367.3803   & 0.2543\\ \hline
HGMCA - 3 & 22.5901 & 2.4231 & 10.9210  & 0.0919\\ \hline
HGMCA - 4 & 23.7845 & 1.6598 & 0.2706  & 0.0844  \\ \hline
\end{tabular}
    \caption{Comparison of RMSE ($\mu K$) and average percent error in the CMB angular power spectrum including and excluding the UT78 mask for WavILC, GMCA\textsuperscript{LASSO}, HGMCA at several levels for skymaps with model 2 dust and synchrotron emissions. GMCA\textsuperscript{LASSO} and HGMCA maps are selected via the input CMB blind criterion described in Section~\ref{sec:ran_seeds}. WavILC provides the construction with the lowest RMSE and $C_\ell$ for high $\ell$, with HGMCA-4 offering a marked improvement at low $\ell$. HGMCA levels 1 and 2 underperform across all metrics due to the smaller scale local variation of model-2. See Figure~\ref{fig:model_2_cl} for a plot across all $\ell$.}
    \label{table:model-2}
\end{table*}

\begin{table*}
    \centering
\begin{tabular}{|l|l|l|l|l|}
\hline
Algorithm   & $|C_{\ell}^{\text{Residual,AME}}|$ & $|C_{\ell}^{\text{Residual,Free-Free}}|$ & $|C_{\ell}^{\text{Residual,Synchrotron}}|$ & $|C_{\ell}^{\text{Residual,Dust}}|$ \\ \hline
GMCA\textsuperscript{LASSO}            & 0.0024          & 0.1519   & 0.0702    &  0.1481 \\ \hline
WavILC            & 0.0006         & 0.0437   & 0.0184    &  0.5459\\ \hline
HGMCA - 1 & 0.0051 & 0.3234 & 0.1474    & 1.1613\\ \hline
HGMCA - 2 & 0.0050 & 0.3165 & 0.1446    & 1.1380\\ \hline
HGMCA - 3 & 0.0009 & 0.0589 & 0.0283  & 0.2141\\ \hline
HGMCA - 4 & 0.0002 & 0.0107 & 0.0056  & 0.0369\\ \hline
\end{tabular}
    \caption{Comparison of correlation of residual CMB map of WavILC, GMCA\textsuperscript{LASSO}, and HGMCA to foreground contaminants. HGMCA levels 1 and 2 continue to underperform due to the small scale local variation of model-2. HGMCA-3 and WavILC have comparable residual contamination while HGMCA-4 improves in every category by at least a factor of three (and a full order of magnitude for dust contamination). See Figure~\ref{fig:model_2_cl_fg} for plots across all $\ell$.}
    \label{table:cor-model-2}
\end{table*}

\subsection{Non-Gaussianity}\label{sec:non-gauss}

The input simulation we described in Section \ref{sec:sim_data} treats the CMB as a Gaussian random field, so a perfect reconstruction would return a map with no non-Gaussianities. In this framework, we cannot fully test HGMCA's ability to construct maps useful for non-Gaussianity studies, but we can probe the amount of non-Gaussianity introduced by our method. In Figure \ref{fig:non_gauss} we compare the change in kurtosis (the fourth standardized moment) between the input CMB maps and the reconstructed map for each algorithm along different latitudes. This comparison was inspired by the analysis done in \cite{L-GMCA}. Within the context of an input map without non-Gaussianity, none of the four algorithms appear to introduce substantial kurtosis from their residuals. Conducting an identical test with skewness rather than kurtosis gives similar results.

\section{Conclusions}\label{sec:conclusion}
We have presented HGMCA, a hierarchical, Bayesian approach to Blind Source Separation for the CMB. The algorithm has been built to tackle one of the principal challenges in precision measurements of ISW signals, gravitational lensing, primordial non-Gaussianity, constraints on isotropy, topological defects, global geometry, etc: separating the CMB from its foregrounds at the map level. Work by \cite{Bobin} has shown that a sparsity-based BSS approach like GMCA can be effective in removing these contaminants. However, vanilla GMCA requires a fixed mixing matrix across the sky and therefore does not have the flexibility to capture the known local variation in the frequency dependence of foregrounds. To address this shortcoming, \cite{L-GMCA} introduced LGMCA. LGMCA subdivides the sky into a series of smaller patches, and independently runs GMCA across each of them. As a result, each patch's frequency dependence is allowed to vary, but it is also blind to the constraining power of the rest of the sky. Further, since the same data must be re-used at different scales of subdivision, the strategy does not admit a generative probabilistic model and requires a heuristic method of choosing between multiple reconstructions.

HGMCA allows for local variation by decomposing the sky into small patches but, unlike LGMCA, it maintains global consistency by enforcing a hierarchy of mixing matrix priors. Further, by dividing the data by scale rather than repeating it at each level of subdivision, HGMCA admits a generative probabilistic model. 

We have provided rigorous testing of HGMCA levels 1, 2, and 3 across simulations generated by \texttt{PySM} at $N_{\rm side} = 256$. Our maps do not vary beam size as a function of frequency and include four different foreground components: dust, synchrotron radiation, free-free emission and anomalous microwave emission.  We have shown that, given the Model 1 \texttt{PySM} templates, HGMCA provides cleaner CMB reconstructions than either GMCA or state-of-the-art Internal Linear Combination-type algorithms, with an order of magnitude less correlation to foreground sources, a $\sim 35\%$ (GMCA) / $\sim 75 \%$ (WavILC) lower cross correlation error for all $\ell$ and $\sim 75\%$/ $\sim 35 \%$ for $\ell>200$, and a $\sim 20\%$ / $\sim 45\%$ reduction in RMSE to the input CMB. When extended to templates with much smaller scale local variation, HGMCA at deeper levels continues to improve over GMCA and performs competitively with WavILC.

Beyond demonstrating that HGMCA incorporates local variation effectively, we have also explored some of the generalization concerns related to sparsity based methods. GMCA and HGMCA require setting hyperparameters which, while theoretically motivated, are set by tuning to simulations. Since even the most advanced simulations will not perfectly reproduce the conditions of the data, it is a natural concern that GMCA and HGMCA may not generalize well to the observed sky. We address this concern by presenting results strictly on maps with different CMB realizations than in our training, and by presenting results on maps with different foreground templates for dust and synchrotron emissions. Further, we show that HGMCA is not particularly sensitive to the exact values of these hyperparameters.

In addition to demonstrating HGMCA's robustness to changes in the CMB, foreground, and hyperparameters, we have also addressed concerns about convergence to a stable CMB reconstruction. The publicly available implementation of GMCA includes a threshold parameter for convergence. In our experience, we found that both this parameter and the order of optimization had a strong effect on the CMB reconstruction even after many iterations. To address this we have introduced GMCA\textsuperscript{LASSO}, and shown that both it and HGMCA converge to a stable solution which depends only minimally on the order of optimization.

Despite these extensive tests, it is still an open question how the algorithm will behave when faced with the additional complexities of higher fidelity simulations. The algorithms on which WavILC is based have been shown to work well even when the data involves effects like varying beam size and CIB contamination, while HGMCA is still untested in these conditions. However, higher resolution will also allow for deeper levels of subdivision, potentially enabling HGMCA to further capitalize on its ability to capture local variations in the foregrounds.

HGMCA's rigorous probabilistic structure, favorable comparison to GMCA and WavILC, and robustness to changes in data suggest it is a well-suited method for use in current and upcoming CMB experiments. As a hierarchical framework, HGMCA also opens the door to incorporating local variation in other BSS-based algorithms with rigorous probabilistic structure. We plan to make the HGMCA code and the CMB reconstruction of the \textit{Planck} dataset available in the near future. Additionally, by using directional spin wavelets we should be able to extend HGMCA to foreground separation on polarization maps (as an example of this, see \cite{SILCspin}). We leave this for future work.

\newpage
\bibliographystyle{mnras}
\bibliography{bibliography} 

\section*{Acknowledgments}
We thank David Alonso, Jason McEwen, Hiranya Peiris, Andrew Pontzen, and Keir Rogers for useful discussions. We would like to thank Benjamin Racine for his help with the simulations.
We also thank Jerome Bobin and Jean-Luc Starck for helpful correspondence.
We especially thank Boris Leistedt for various contributions to this project.
CD was supported by NSF grant AST-1813694 and Department of Energy (DOE) grant DE-SC0020223. CD was also supported by the Star Family Challenge for Promising Scientific Research and the Dean's Competitive Fund for Promising Scholarship at Harvard University. MH was supported through NSF Award DGE-1650112. SWC was supported through the KIPAC-Chabolla fellowship.

\newpage
\appendix
\section{CMB Source Separation Techniques}\label{app:rw}
Here we highlight the four models used most recently by the \textit{Planck} collaboration, SEVEM, SMICA, Commander, and NILC \citep{SEVEM,SMICAorig,Commander2,NILC}, as well as providing some general background on template fitting and internal linear combination.

\subsection{Template Fitting}
In template fitting, Equation~\eqref{eq:BSS} is simplified to
\begin{equation}\label{eq:Template}
d = \tilde{d} + \alpha t,
\end{equation}
where the sky $d$ is a linear combination of the CMB $\tilde{d}$, and a template $t$ mapping anisotropies \citep{Land}. To isolate the CMB, one must construct the template $t$ and solve for $\alpha$. A standard method is to solve for the maximum likelihood value of $\alpha$ by minimizing $\chi^2$. This technique can be extended to add multiple templates for different foreground components \citep{SEVEM, Jaffe}.

Template fitting techniques can be either parametric (using external information to parametrize the components of the sky) or blind. Parametric techniques develop templates iteratively, comparing results with external data and theory to develop new templates.

\subsection{Internal Linear Combination}\label{app:ILC}
Internal Linear Combination (ILC) is another technique that focuses solely on recovering the CMB, but relies on the assumption that the CMB signal is independent of frequency rather than using foreground templates. ILC observes that for any affine combination\footnote{An affine combination is a linear combination with coefficients that sum to $1$} of input maps $X_i$:
\begin{equation}\label{eq:ilc}
    \sum w_iX_i = C + \sum_i w_i ( f_i + n_i),
\end{equation}
where $C$ is the CMB, $f_i$ is foreground signal, and $n_i$ is noise. Under the assumption that the CMB is independent from the foreground and noise, it is then possible to minimize the foreground contamination by finding the set of weights $w_i$ that minimizes the variance of $\sum w_iX_i$, effectively setting the rightmost term of Equation \eqref{eq:ilc} to $0$. A derivation of the formula used to calculate the weights can be found in \cite{ILC}. 

ILC's only major assumption is the frequency dependence of the CMB. It can operate in pixel, harmonic, or wavelet space, and can be solved for analytically. The reconstruction quality and noise properties are also well-understood theoretically \citep{ILC}. However, it does not tap into the extra information that can be gained by making more advanced assumptions about the foregrounds, such as their sparsity in the wavelet basis, which can enable more accurate reconstructions.

\subsection{SEVEM}

SEVEM is an internal template fitting technique currently used by the \textit{Planck} collaboration. SEVEM creates templates without external data by subtracting \textit{Planck} maps at close frequencies. Since the CMB is the same across frequencies, this isolates foreground components. SEVEM's template fitting is done in real space at every position in the sky. Given $n$ such templates $t_i$, a position $x$ and frequency $\nu$ the CMB is then given by the difference from the original map $d(x,\nu)$ and a linear combination of templates:
\[
T_{CMB}(x,\nu) = d(x,\nu) - \sum\limits_{i=1}^n \alpha_{i}(\nu) t_{i}(\nu)
.\]
No assumptions on foreground or noise are used in SEVEM, making it a robust technique \citep{SEVEM}. \cite{planck2015foregrounds} suggests SEVEM is best used to deal with bright point sources in polarization, and notes that for a number of frequencies, it is the only technique that can provide independent CMB reconstructions. 

\subsection{NILC}\label{app:NILC}
ILC can be performed either in harmonic or pixel space, but fails to take into account that noise and foreground dominate at different angular frequencies in harmonic space, and different Galactic latitudes in pixel space \citep{NILC2}. In order to deal with this, \cite{NILC} introduced Needlet Internal Linear Combination (NILC), an extension of ILC to wavelet space. NILC uses a specific type of wavelet known as the needlet \citep{needlet} which avoids ILC's issue by having compact support in the harmonic domain along with localization in the pixel domain \citep{NILC2}.
More recently, \cite{SILC} proposed Scale-discretized, directional wavelet ILC (SILC). By using non-axisymmetric wavelets, SILC takes advantage of orientation in signal structure to further fine tune the weights of ILC. \cite{planck2015foregrounds} note that NILC works well at high $\ell$, with performance comparable to SMICA. \\

The covariance matrix used for NILC and SILC (as well as WavILC in this work) is empirically calculated by convolution with a Gaussian beam. First an approximate covariance matrix $C_{ij}^\text{approx}$ is calculated by:
\begin{align}
    C_{ijk}^{\mu,\text{approx}} &= x_{ik}^\mu x_{jk}^\mu
\end{align}
where $i$ and $j$ are two frequency bands, $k$ is a pixel coordinate, and $\mu$ is the wavelet scale. The convolution is then calculated in the harmonic space as:
\begin{align}
    \tilde{C}_{i j \ell m}^\mu &= \sum_{\ell =0}^{2 \ell_\text{max}} \sum_{m=-\ell}^{\ell} g_\ell \tilde{C}_{i j \ell m}^{\mu,\text{approx}}
\end{align}
The final weights are then calculated as:
\begin{align}
    w_{ik}^\mu = \frac{\sum_{j = 1}^{N_\text{freq}} C_{i j k}^\mu e_j }{
    \sum_{j' = 1} \sum_{j = 1}^{N_\text{freq}} e_{j'} C_{j' j k}^\mu e_j }
\end{align}
where $e$ is a vector of ones.

\subsection{SMICA}

Independent Component Analysis (ICA) is a popular BSS method in the machine learning community \citep{ICA}. ICA relies on two major assumptions: that the sources we wish to separate are both independent and non-Gaussian. Let $W$ be the psuedo-inverse of $A$. Then a single source $s_i$ may be expressed as
\begin{align}
    s_i = w_iX.
\end{align}
The basic concept of ICA is to compute $w_i$, and hence $s_i$, by maximizing the non-Gaussianity of $w_iX$ via measures such as kurtosis or negentropy. Further, since sources are assumed to be uncorrelated, optimizing over all sources is not difficult \citep{ICA}. 

While the foreground elements and CMB may be independent, the CMB is Gaussian. To avoid this issue, \cite{SMICAorig} introduced a modification of the method called spectral matching ICA (SMICA). SMICA fixes the issue of the CMB's gaussianity by trading non-Gaussianity for spectral distinction between sources. SMICA follows essentially the approach of ICA described above, but instead computes $S$ and $A$ via maximizing a spectral matching criterion exhibited by the CMB and foreground sources \citep{SMICA}.

As initially proposed, SMICA analyzed spectral matching in the Fourier basis. This method is sub-optimal for capturing non-stationary foreground contaminants, even when using localized spectral statistics \citep{SMICAmid}. \cite{wavelets} propose solving this issue by using a wavelet rather than Fourier basis. Wavelets are better suited to deal with non-stationary components due to their locality in both pixel and harmonic space. \cite{planck2015foregrounds} suggest using SMICA for mid to high $\ell$.
\subsection{Commander}
Commander is a Bayesian, forward parametric technique with the goal of computing the joint posterior distribution \citep{Sampling}, $P(A,S|X)$. Commander is realized via Gibbs sampling, a statistical method for sampling joint distributions from iteratively sampling conditional distributions of $A$ and $S$. In particular, $A$ and $S$ are updated by repeatedly sampling from:
\begin{align}
A &\sim P(A|S,X),\\
S &\sim P(S|A,X).
\end{align}
Once the Gibbs sampling has converged and the posterior distribution appropriately sampled, one can compute desired empirical statistics for $A$ and $S$, such as the mean and standard deviation maps for each component. 

Originally, this technique only succeeded in producing low resolution maps, but recent incarnations such as Commander-Ruler \citep{planck2013cosmo} and Commander 2 \citep{planck2018cosmo} fix this issue. \cite{planck2015foregrounds} and \cite{planck2018foregrounds} recommend using the Commander framework over other techniques at low $\ell$ due to its lower ``internal tension between low, intermediate, and high multipoles".

\section{Scale-discretized wavelet transform and patch decomposition}\label{app:wavelet}

The wavelet transform of a square integrable signal $f$ on the two-dimensional sphere is a decomposition into wavelet coefficient maps $W^{\Psi^{j}}(\omega)$, themselves signals on the sphere, indexed with the wavelet scale $j$ and calculated as 
\begin{align}
  W^{\Psi^{j}}(\omega)   =  \int_{S^2} {\rm d}\Omega(\omega^\prime) f(\omega^\prime) ( \mathcal{R}_\omega \Psi^{j} )^*(\omega^\prime).
\label{wav1}
\end{align}
Reconstruction of the input signal $f$ is achieved through
\begin{align}
	\quad f(\omega) = \sum_{j}  \int_{S^2} {\rm d}\Omega(\omega^\prime) W^{\Psi^{j}}(\omega^\prime)(\mathcal{R}_{\omega^\prime} \Psi^{j})(\omega).
\end{align}
Those operations are interpreted as convolutions on the sphere, relying on a rotation operator on the sphere $\mathcal{R}_\omega$.
The transform is fully characterized by the choice of set of wavelets $\Psi^{j}(\omega)$, which are also signals on the sphere, typically compact and localized on the north pole. Some desirable properties of the transform include 1) invertibility, which amounts to the decomposition and reconstruction operations being lossless, 2) good localization properties of the wavelets in both pixel and harmonic space, 3) the computational cost of performing the convolutions above. 

We adopt the framework of scale-discretized wavelets \citep{s2let1, s2let2,wiaux2008exact}, as implemented in the \texttt{S2LET} code\footnote{\url{https://github.com/astro-informatics/s2let}}.
Central to this approach is the use of the \texttt{SSHT} sampling theorem \citep{ssht}, to convert signals on the sphere (and stored on the \texttt{SSHT} pixelization) into spherical harmonic coefficients, and conversely, in a theoretically exact fashion (\textit{i.e.}, lossless for bandlimited signals). 
This can be exploited to perform the convolutions in harmonic space, which is fast and exact since it reduces to a discrete sum over harmonic coefficients.
As a result, the wavelet transform itself is lossless and invertible, which in practice leads to errors of the order of floating point accuracy when transforming a signal back and forth.
Furthermore, scale-discretized wavelets are based on wavelets $\Psi^{j}(\omega)$ that are simply defined in harmonic space and exactly compact, which leads to further computational improvements over existing approaches making use of wavelets with broad harmonic support \citep{s2let1, s2let2, 2015arXiv150906767M}.

Note that in the concise presentation above we have omitted the scaling function, which is akin to an extra wavelet but captures the low-frequency, low-resolution information of the signal, typically not included in $\Psi^{j}(\omega)$.
A full presentation can be found in \citep{s2let2}.
We also focused on axisymmetric wavelets, but scale-discretized wavelet can be made directional  \citep{s2let1}.

The main parameters of the wavelet transform are the band limit of the signals (for all spherical harmonic transform operations) and the details of the wavelets. 
Axisymmetric scale-discretized wavelets are characterized by a parameter $\lambda$. Each wavelet $\Psi^{j}$ has a compact harmonic-space support in $[\lambda^{j-1}, \lambda^{j+1}]$, with the peak at $\lambda^{j}$. For the results in this work we used $\lambda=3$, a bandlimit of $\ell=621$, and a minimum $J_0 = 1$.

In our BSS formalism, the linearity of the wavelet transform allows us to write the signal $X$ as a matrix product $AS$ in the wavelet basis, just like we would do in pixel or harmonic space. 
The elements of $X$ and the rows of the source matrix $S$ would be wavelet coefficients, pixels, or harmonic coefficients, respectively. LGMCA and HGMCA use a decomposition of the map $X$ into patches. This decomposition is done in the wavelet space (i.e. where X and S are wavelet coefficients) to avoid the computational cost of repeatedly conducting a forward and inverse wavelet transform. Because the patches are never allowed to probe scales near the wavelet support, these patches should have a close correspondence to the same patch in the original space. This decomposition also has the added benefit of smoothing out the harsh boundaries that would occur in a pixel-based decomposition.
\section{Non-negativity Constraints for GMCA}\label{ngmca}

For applications of GMCA, LGMCA, and HGMCA to CMB foreground separation our mixing matrix $A$ cannot be negative. It is a representation of the contribution of a given source at a given frequency, and cosmological sources cannot contribute negative flux. With this in mind, \cite{nGMCA} introduce a non-negative version of GMCA. Applying a prior to each column of $A$ which is uniform over the positive orthant of the unit sphere\footnote{This also accounts for the L2-normalization of $A$, and thus provides a Bayesian interpretation of \citep{nGMCA}}, the optimization problem becomes
\begin{equation}
\text{argmin}_{A,S} \ \lambda_S||S||_0 + ||X - AS||_F^2 + i^+(A),
\end{equation}
where $i^+(A)$ is the characteristic function of the positive orthant,
\begin{equation}
i^+(A)=
\begin{cases}
0 & A \geq 0\\
\infty & A < 0.
\end{cases}
\end{equation}

\cite{nGMCA} present several optimization techniques for this problem, and empirically show that for the case of the sky where there are few sources to separate, it is the naive algorithm of setting any non-negative values to $0$ which performs best. This same naive modification can be made to both LGMCA and HGMCA to enforce non-negativity.

\section{HGMCA Update Rules}\label{app:HGMCA}

We start by writing out the full loss function for HGMCA,
\begin{align}
    \mathcal{L} &= \underbrace{\sum_l  \sum_{\mu_l} \sum_{p \in \{p_l\}} i^+(A_p^l) + \lambda_S||S_{p}^{l,\mu_l}||_1
    + ||X_{p}^{l,\mu_l} 
    -A_{p}^{l}S_{p}^{l,\mu_l}||^2_F}_{\text{source prior and reconstruction error}} \nonumber\\
    &+ \underbrace{\sum_l \sum_{p\in\{p_{l-1}\}} \sum_{i \in 1, \dots 4}
    \lambda_A ||A_{p}^{l}-A_{p,i}^{l}||^2_F.}_{\text{A matrix dependence across levels}.}
\end{align}

The notation $l$ denotes the level, $\mu_l$ denotes the wavelet scales with maximum subdivision level $l$, and $p\in \{p_l\}$ indicates the patches at subdivision level $l$ (for example $p \in \{p_2\}$ are $p=p_1,p_2$ with $p_1,p_2 \in 1,2,3,4$). The norm $||M||^2_F$ is the Frobenius or 2-norm and $||M||_1$ is the 1-norm of a matrix $M$.

Since all the wavelet scales at a single level share the same mixing matrix, we concatenate the different $\mu_l$ scales together so that the source matrix $S^l_{p}$ at subdivision level $l$ on a patch $p \in \{p_l\}$. Therefore, $S^l_{p}$ now has dimensions of $N_\mathrm{S}$ by (number of pixels times number of scales in $\mu_l$), allowing us to drop the wavelet superscript. Similarly, for each level the dimensions of the data matrix $X^l_p$ will be number of frequencies by (number of pixels times number of scales in $\mu_l$).

We are interested in knowing the update rule for a specific row of a source matrix $S^l_p$ or column of a mixing matrix $A^l_p$. We can therefore also drop the subscripts and superscripts related to the level and particular patches for simplicity. In the following, we denote $s{_i}$ as the $i^{\text{th}}$ row of $S^l_p$ corresponding to source $i$ and $a^i$ as the $i^{\text{th}}$ column of the mixing matrix $A^l_p$ on a patch $p$ corresponding to source $i$. Dropping all the terms that have no dependence on $s_i$ we obtain the following loss function,

\begin{align}
\mathcal{L}_{s_i} &= \lambda_S ||s_i||^1 + ||X-\sum_j a^j s_j ||^2_F,  
\end{align}
which has contributions from our sparsity prior and the probability of generating $X^l_p$ from $A^l_p$ and $S^l_p$.
For all $j \neq i$ the value of $s_j$ is fixed, so we can further simplify the above loss function by introducing a new term $R_i = X - \sum_{j\neq i} a^js_j$,
\begin{align}
\mathcal{L}_{s_i} &= \lambda_S ||s_i||^1 + ||R_i- a^i s_i ||^2_F  \nonumber\\
&= \lambda_S ||s_i||^1 + \Tr \left[\left(R_i- a^i s_i\right)^T\left(R_i- a^i s_i\right) \right] \nonumber\\
&= \lambda_S ||s_i||^1 + \Tr \left[R_i^T R_i - R_i^T a^i s_i - s_i^T (a^i)^T R_i + s_i^T(a^i)^T a^i s_i \right] \nonumber\\
&= \lambda_S ||s_i||^1 + \Tr \left[s_i^T(a^i)^T a^i s_i - 2 s_i^T (a^i)^T R_i \right],
\end{align}
where we have dropped the $R_i^T R_i$ since it has no $s_i$ dependence and taken advantage of the invariance of trace under transpositions.

We can now proceed to find the parameter $s_i$ that minimizes the loss function,
\begin{align}
0 = \frac{\partial \mathcal{L}_{s_i}}{\partial s_i} &= \frac{\partial}{\partial s_i} \lambda_S ||s_i||^1 + 2 (a^i)^T a^i s_i - 2 (a^i)^T R_i .
\end{align}
Note that $(a^i)^T a^i$ is simply the square of the 2-norm of the $i^\text{th}$ column of $A$. Due to the degeneracy between scaling a column of $A$ and a row of $S$ by inverse factors it is common practice in the BSS literature to set the 2-norm of $A$ to 1. With this additional simplification we are left with
\begin{align}
s_i &= (a^i)^T R_i - \frac{\partial}{\partial s_i} \frac{\lambda_S}{2} ||s_i||^1.
\end{align}
We solve this equation through the LASSO shooting algorithm, which gives the following solution in matrix form \citep{LASSO},

\begin{align}
s_i &= \begin{cases}
(a^i)^T R_i - \frac{\lambda_S}{2} \text{sign} ((a^i)^T R_i) & |(a^i)^T R_i|> \frac{\lambda_S}{2} \\
0 & |(a^i)^T R_i| \leq \frac{\lambda_S}{2}
\end{cases},
\end{align}
with sign being an element-wise sign function that returns the sign of the number or 0 if the number is 0. 

We then have to obtain an analogous optimization equation for $a^i$. We use $\tilde{A}$ to denote the mixing matrix that defines the distribution from which $A$ is drawn. As an example, for $A = A^1_{p_1}$ this would be $\tilde{A}=A^0$. We use $B_j$ to denote the $j^\text{th}$ matrix that is drawn from $A$. As an example, for $A = A^1_{p_1}$ this would be $B_j=A^2_{p_1,j}$ for $j \in 1,2,3,4$. We can now write the full loss function dropping all terms that have no dependence on $a^i$. Additionally, we add a Lagrange multiplier term $\lambda_m$ that enforces that the two-norm of a column of A is 1,

\begin{align}
\mathcal{L}_{a^i} &= ||R_i- a^i s_i ||^2_F + \lambda_A ||A - \tilde{A}||^2_F + \sum_j \lambda_A ||A - B_j||^2_F \nonumber\\
&+ \lambda_m \left(||A||_F^2-1\right) \nonumber\\
&=  \Tr \left[(a^i)^T a^i s_i s_i^T - 2 (a^i)^T R_i s_i^T \right]\\
&+ \lambda_A \Tr \left[(a^i)^T a^i - 2 (a^i)^T \tilde{a}^i \right]  \nonumber\\
&+ \sum_j \lambda_A \Tr \left[(a^i)^T a^i - 2 (a^i)^T b^i_j \right] + \lambda_m \Tr \left[(a^i)^T a^i\right].
\end{align}
This includes terms from the mixing matrix hierarchy prior as well the generative distribution of $X^l_p$. We can then find the parameter $a^i$ that minimizes the loss function,

\begin{align}
\frac{\partial \mathcal{L}_{a^i}}{\partial a^i} &= 0 = 2 a^i ||s_i||^2 - 2 R_i s_i^T + 2 \lambda_A \left( (1+N_j)a^i - \tilde{a}^i - \sum_j b^i_j \right) \nonumber \\
&+ 2 \lambda_m a^i \\
a^i &= \frac{1}{||s_i||^2 + \lambda_A(1+N_j) - \lambda_m} \left[R_is_i^T + \lambda_A \tilde{a}^i + \lambda_A \sum_j b^i_j \right],
\end{align}
where $N_j$ is the number of matrices drawn from the distribution defined by $A$. Note that this is the most general solution for a column of the mixing matrix. We can then solve for $\lambda_m$ by plugging in $\frac{\partial \mathcal{L}}{\partial \lambda_m} = 0 = \Tr[(a^i)^Ta^i]-1$, which yields

\begin{align}
(a^i)^T \frac{\partial \mathcal{L}_{a^i}}{\partial a^i} &= 0 = 2 ||s_i||^2 - 2 (a^i)^T R_i s_i^T + 2 \lambda_m \nonumber\\
&+ 2 \lambda_A \left( (1+N_j) - (a^i)^T \tilde{a}^i - (a^i)^T \sum_j b^i_j \right)  \\
\lambda_m &= -||s_i||^2 + (a^i)^T R_i s_i^T \nonumber \\
&- \lambda_A \left( (1+N_j) - (a^i)^T \tilde{a}^i - (a^i)^T \sum_j b^i_j \right).
\end{align}
Let $[x]_+$ denote $max(0,x)$, then plugging this back into our original equation and enforcing non-negativity we obtain 
\begin{align}
a^i &= \left [\frac{R_i s_i^T + \lambda_A \tilde{a}^i + \lambda_A \sum_j b^i_j}{||R_is_i^T + \lambda_A \tilde{a}^i + \lambda_A \sum_j b^i_j||_{F}}  \right ]_+,
\end{align}
which is the vector divided by its 2-norm.

It is possible that there are no wavelet scales at the level corresponding to $A$ in which case the solution would look like
\begin{align}
a^i &= \left [\frac{1}{|| \tilde{a}^i + \sum_j b^i_j||_{F}} \left[ \tilde{a}^i + \sum_j b^i_j \right] \right]_+,
\end{align}
which is the intuitive form we would expect (in the graphical model, an average of all the matrices it shares edges with). We can similarly deal with the case where there are no $\tilde{A}$ or $B_j^i$ by dropping the corresponding terms.

\section{GMCA and HGMCA parameters}\label{app:params}
Table \ref{table:all_params} describes the parameters used for the GMCA and HGMCA configurations presented in Section \ref{sec:results}. Map Normalization refers to an extra pre-processing step where either nothing is done (None) or the region contained within the UT78 mask is scaled by $0.5$ (Center). The overall effect on the results is small, but HGMCA appears to perform slightly better with this pre-processing. The parameters of all three algorithms are optimized to minimize RMSE on a set of training simulations. All the results presented in this work are from separate simulations where either the CMB realization is modified (Section \ref{sec:alg_comp}) or all the source realizations are modified (Section \ref{sec:for_temp})
\begin{table}
    \centering
\begin{tabular}{lll}
\hline
Algorithm         & GMCA      & HGMCA-1/2/3                  \\ \hline
$\lambda_S$          & 50                     & 50                 \\ \hline
$N_\mathrm{S}$ & 5         & 5                                \\ \hline
$\lambda_{\rm CMB}$ & $10^{11}$ & $10^{11}$   \\ \hline
Map Normalization & None      & Center                      \\ \hline
$\lambda_A$          & N/A       & $5 \times 10^{10}$  \\ \hline
\end{tabular}
    \caption{The parameters for best performing GMCA and HGMCA algorithms. As discussed in Section \ref{sec:generalization}, these parameters are chosen with an initial set of simulations by testing which values return the lowest RMSE error. The final results presented throughout the rest of the paper are from a different set of simulations to show that we have not overfit to the initial simulation set.}
    \label{table:all_params}
\end{table}

\section{GMCA Optimization Strategies}\label{app:GMCA_Comparison}

\begin{figure}
    \centering
    \includegraphics[scale=0.35]{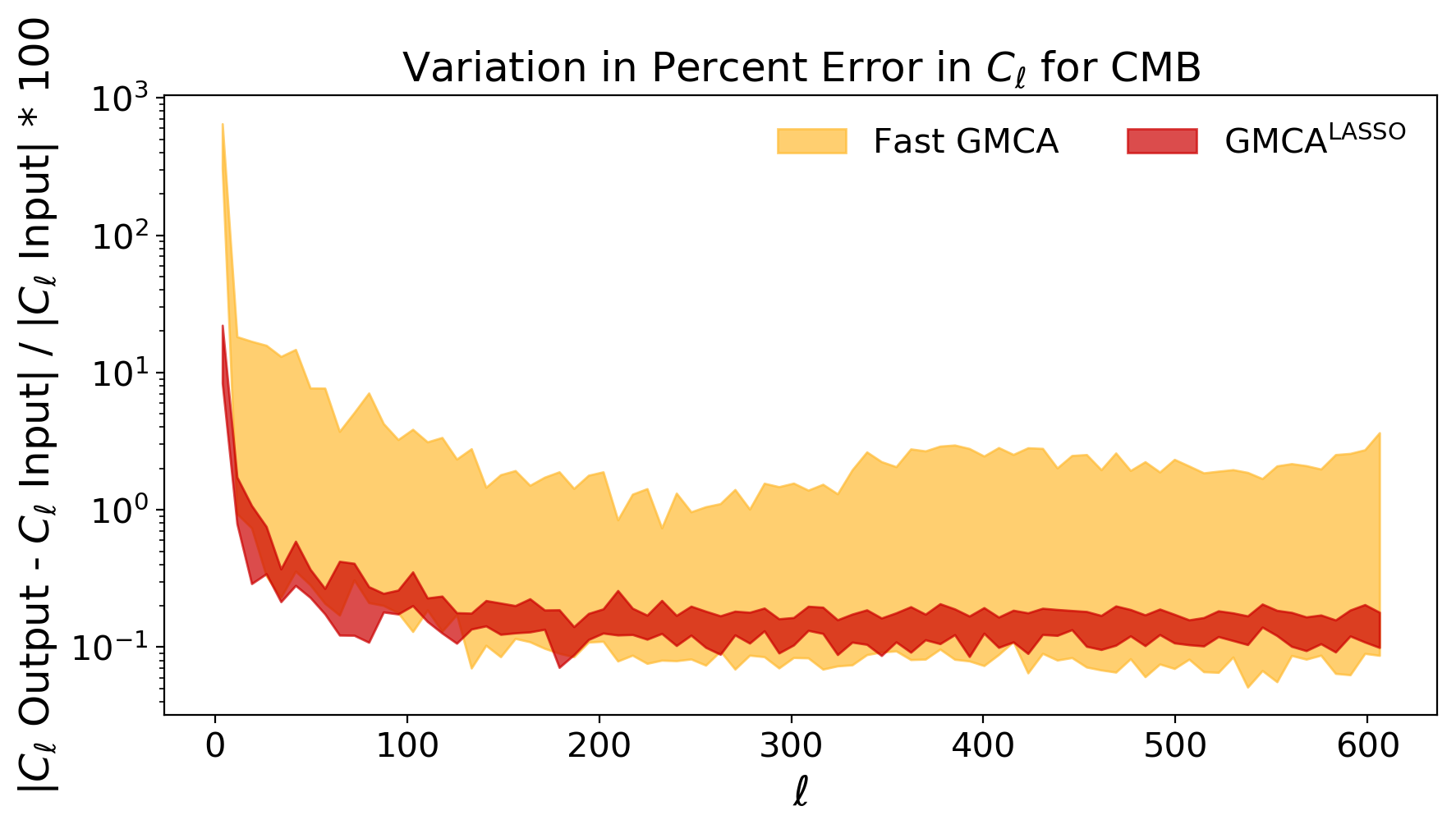}
    \caption{The variation in CMB $C_\ell$ error due to changes in the random seed for Fast GMCA and GMCA$^{\text{LASSO}}$. For both algorithms, the hyperparameters are fixed at an optimum value determined by the methodology described in Section \ref{sec:generalization}. While both produce competitive best-case results, Fast GMCA has significantly larger variance.\label{fig:fastlasso}}
\end{figure}

To test the performance and stability of Fast GMCA versus GMCA$^{\text{LASSO}}$, we follow the parameter selection criteria of section \ref{sec:generalization} and test each algorithm across a new instance of the CMB and across multiple random seeds. Our Fast GMCA implementation follows \cite{Bobin}. This includes using hard-thresholding and manually enforcing that the CMB column match theoretical expectations. Figure~\ref{fig:fastlasso} shows the best and worst performances of Fast GMCA and GMCA$^{\text{LASSO}}$ across these seeds. The results show that while Fast GMCA's performance across $C_{\ell}$'s is competitive with GMCA$^{\text{LASSO}}$, its optimization is significantly more sensitive to random initialization. In particular, the error in $C_{\ell}'s$ from Fast GMCA varies by nearly two orders of magnitude, whereas the error in GMCA$^{\text{LASSO}}$ is within about a factor of $2$ across all multipoles. The same results emerge if the CMB is perturbed and the random seed is held fixed.

\section{Reconstructed Maps}
In this section, we present the reconstructed CMB maps from WavILC, GMCA, HGMCA-1,HGMCA-2, and HGMCA-3 corresponding to the results presented in Section~\ref{sec:results}.
\begin{figure*}
    \centering
    \includegraphics[scale=0.35]{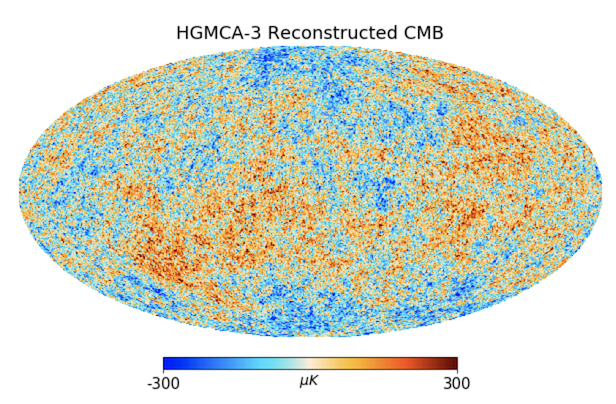}
    \includegraphics[scale=0.35]{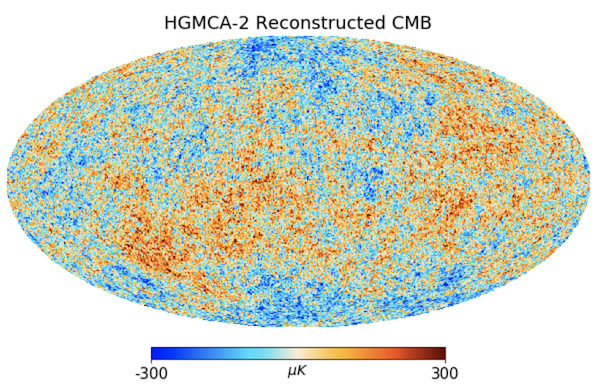}
    \includegraphics[scale=0.35]{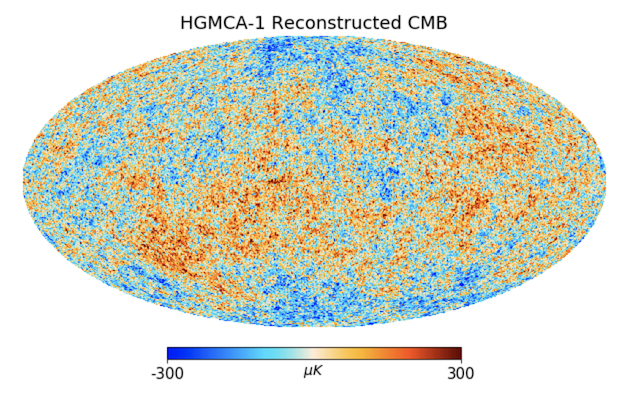}
    \includegraphics[scale=0.35]{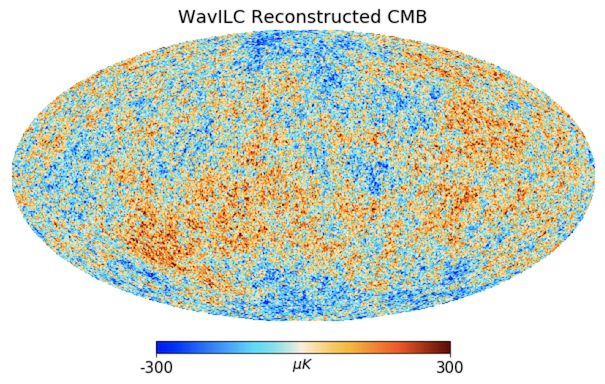}
    \includegraphics[scale=0.35]{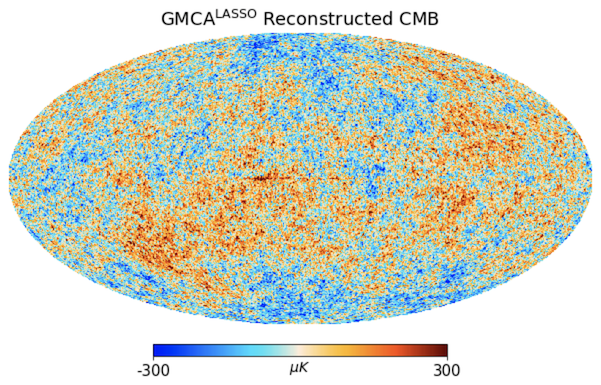}
    \caption{From top left to bottom: the reconstructed CMB sky for HGMCA-3, HGMCA-2, HGMCA-1, WavILC, and GMCA\textsuperscript{LASSO}.}
    \label{fig:CMB_maps}
\end{figure*}
\section{Figures for Template Model $2$}\label{app:gen}
\begin{figure*}
    \centering
    \includegraphics[scale=0.36]{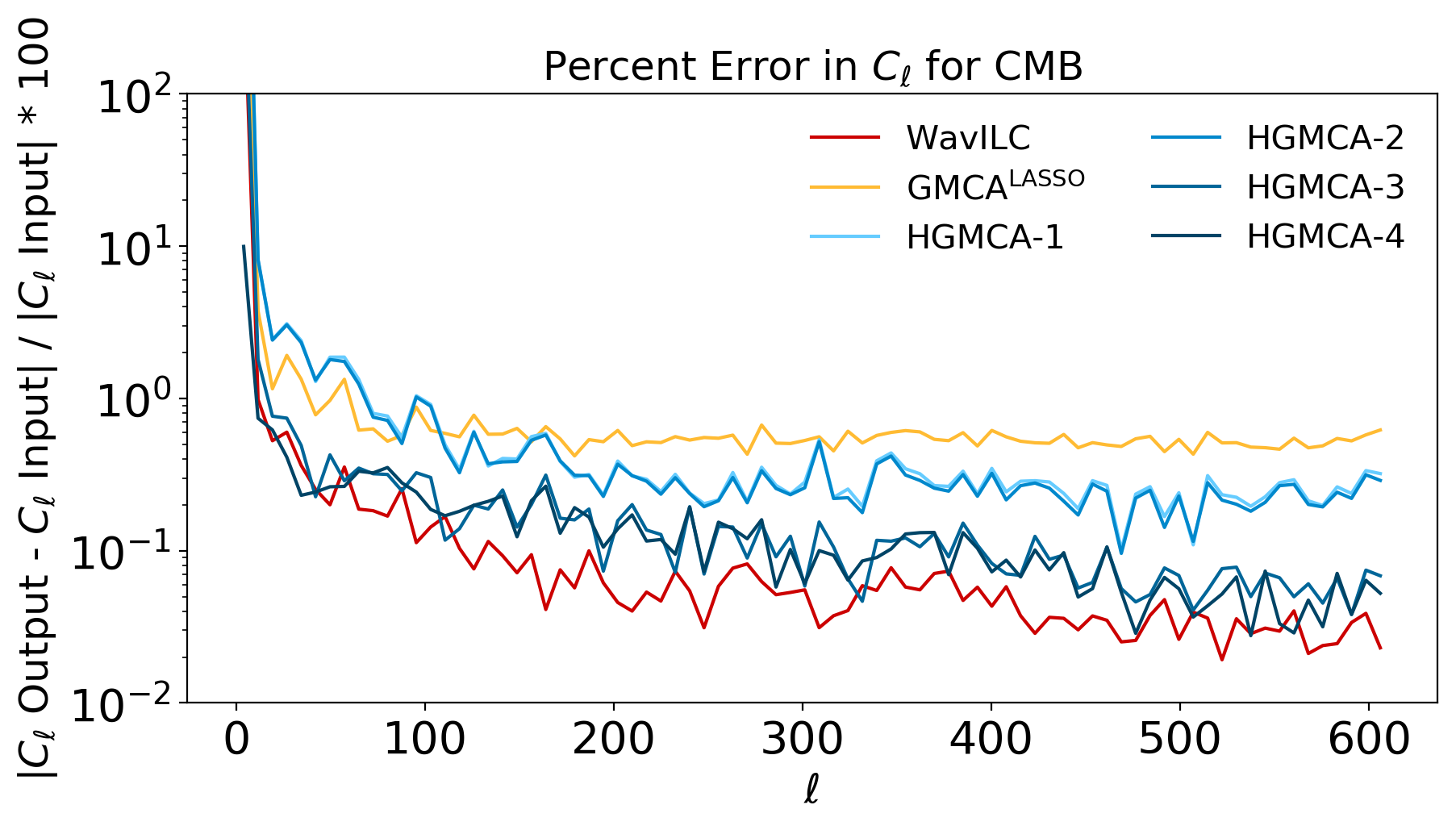}
    \caption{For the generalization test described in Section \ref{sec:for_temp}, percent error in $C_\ell$ as a function of multipole moment $\ell$ for each of the algorithms.}
    \label{fig:model_2_cl}
\end{figure*}

\begin{figure*}
    \centering
    \includegraphics[scale=0.35]{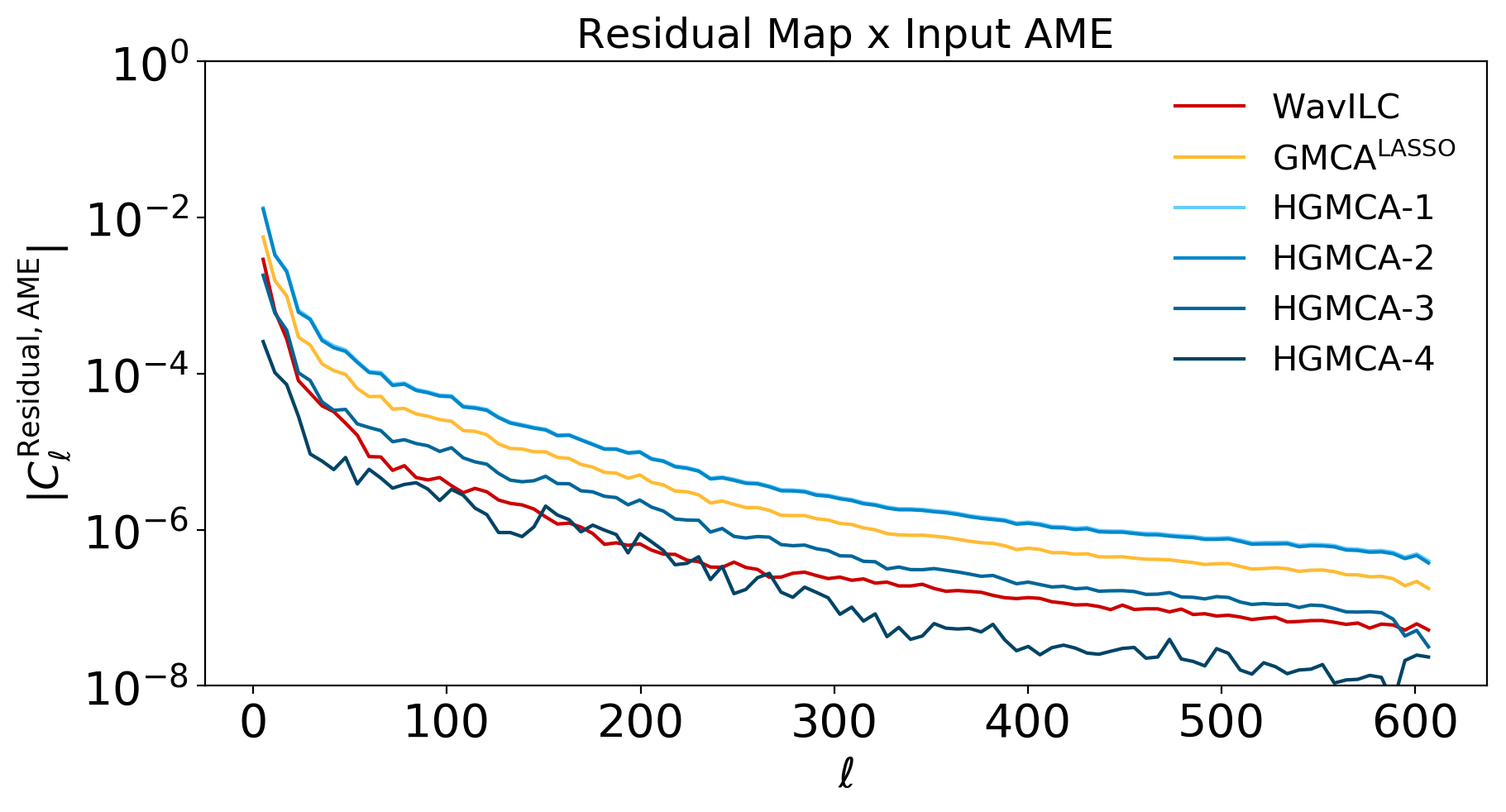}
    \includegraphics[scale=0.35]{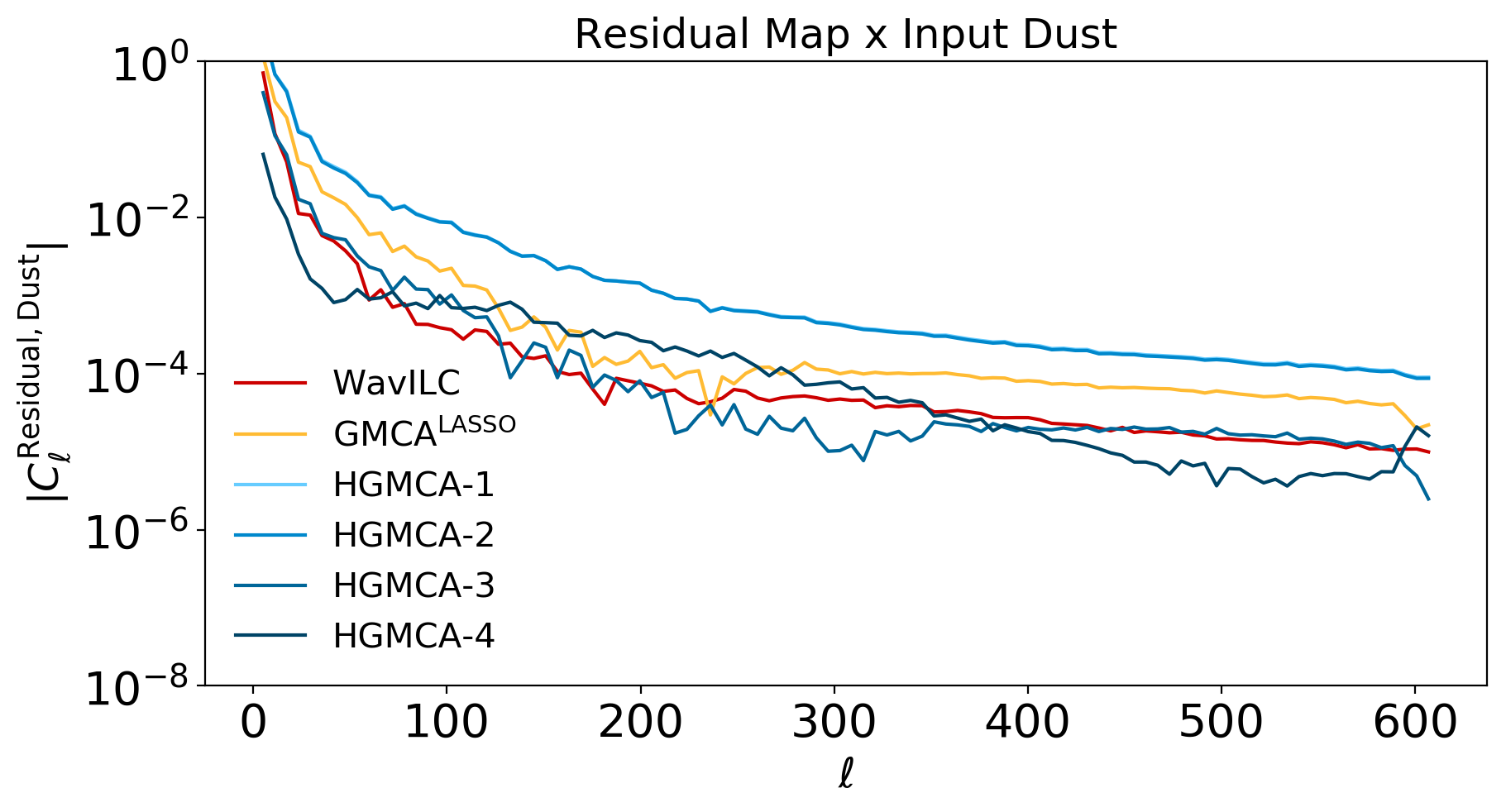}
    \includegraphics[scale=0.35]{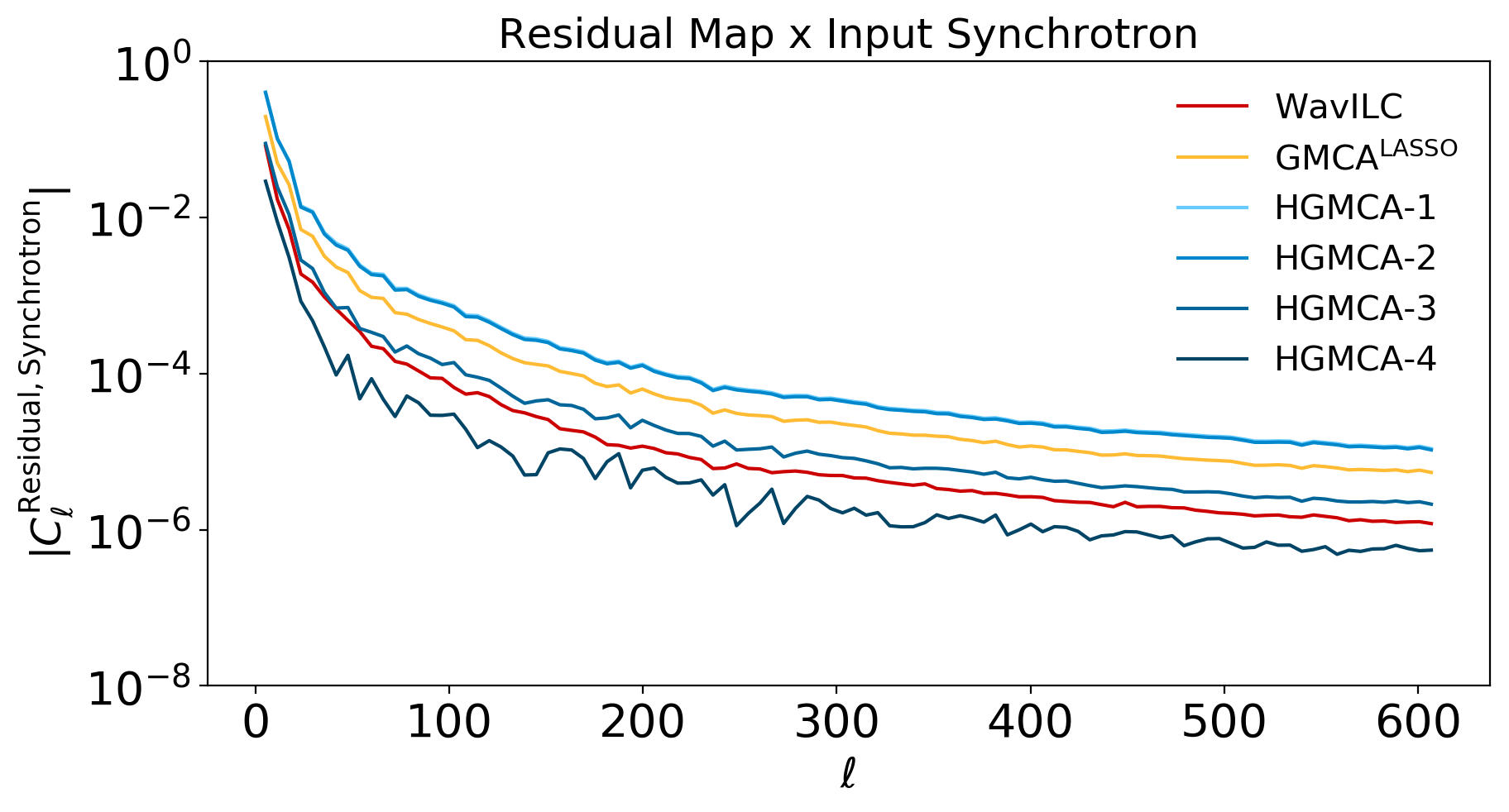}
    \includegraphics[scale=0.35]{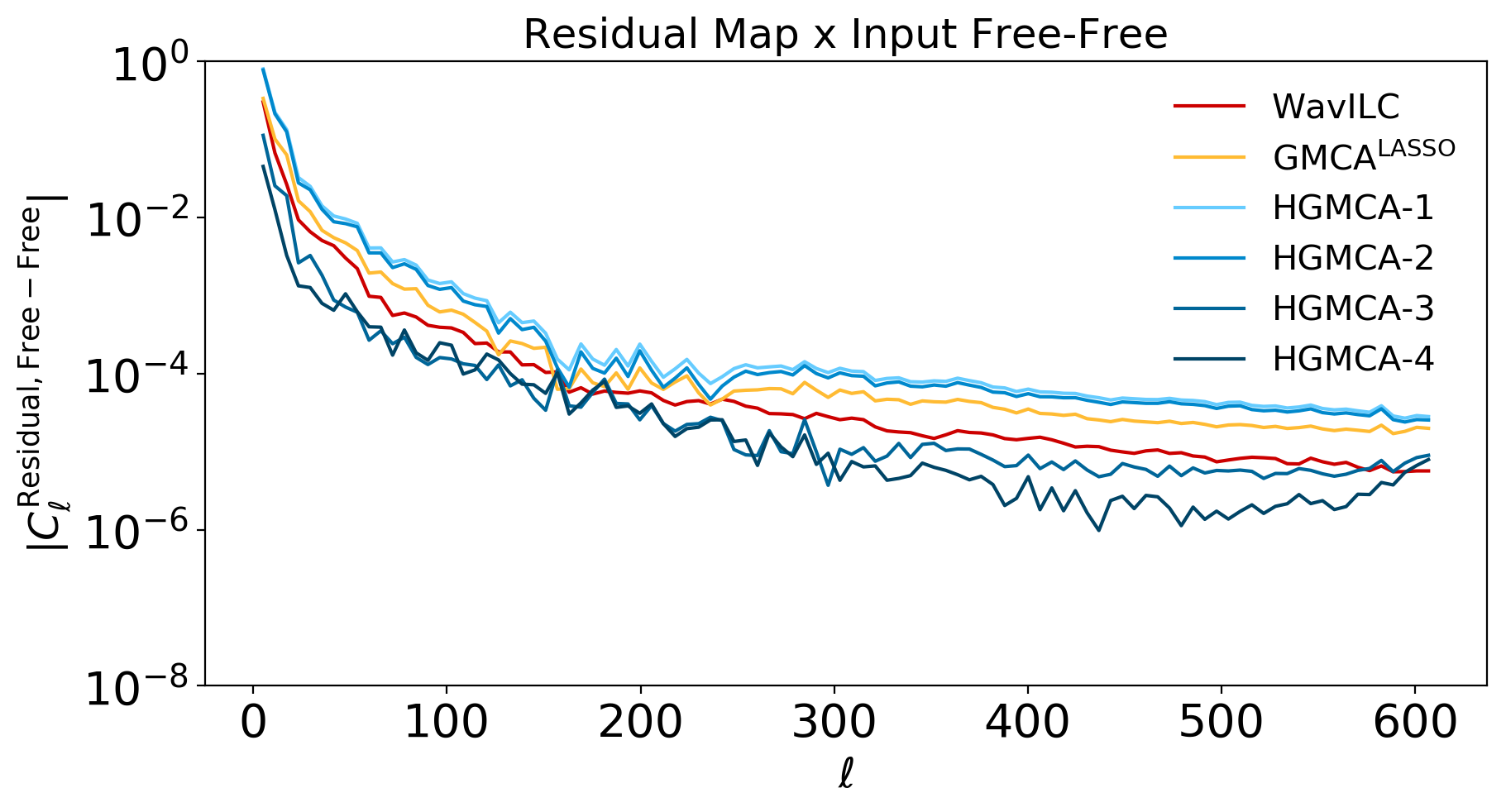}
    \caption{For the generalization test described in Section \ref{sec:for_temp}, magnitude of the cross-spectra between the residual CMB for each algorithm and the input AME, dust, synchrotron, and free-free foreground contaminants.}
    \label{fig:model_2_cl_fg}
\end{figure*}

In this section, we present the plots for the generalization test in Section \ref{sec:for_temp}. Specifically, we show the percent error in the CMB $C_\ell$'s in Figure \ref{fig:model_2_cl} and the magnitude of the cross spectra between the residual CMB and the foregrounds in Figure \ref{fig:model_2_cl_fg}.

\section{Propagated Foreground Residuals and Source Degeneracy}
\label{app:prop_resd}

\begin{figure*}
    \centering
    \includegraphics[scale=0.35]{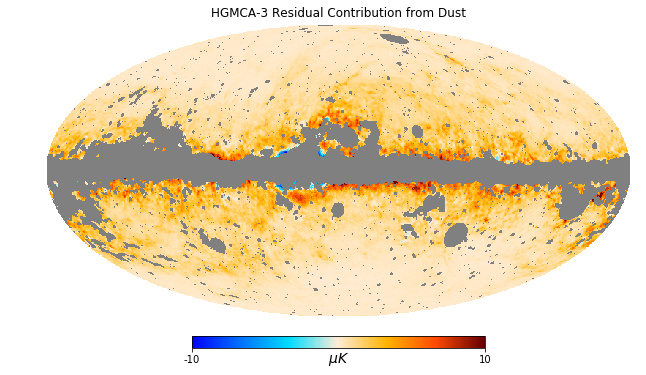}
    \includegraphics[scale=0.35]{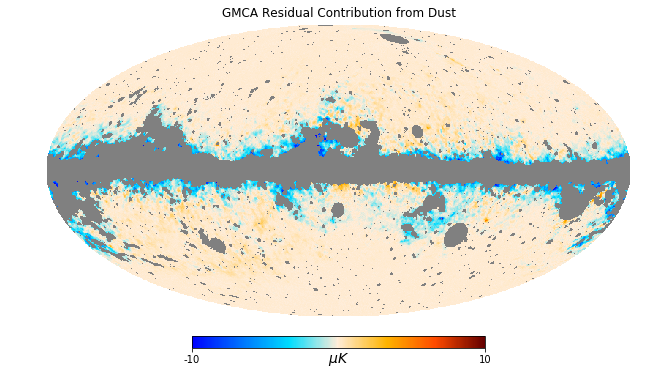}
    \includegraphics[scale=0.35]{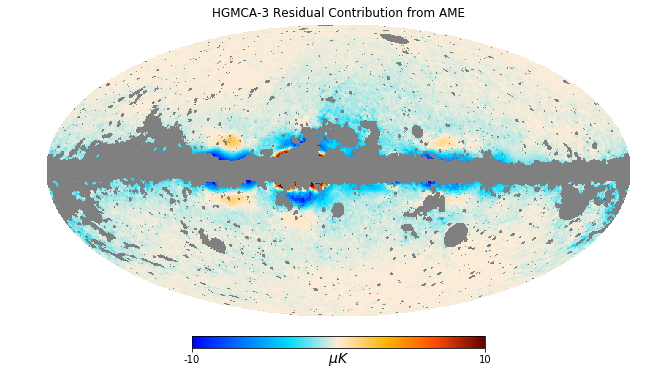}
    \includegraphics[scale=0.35]{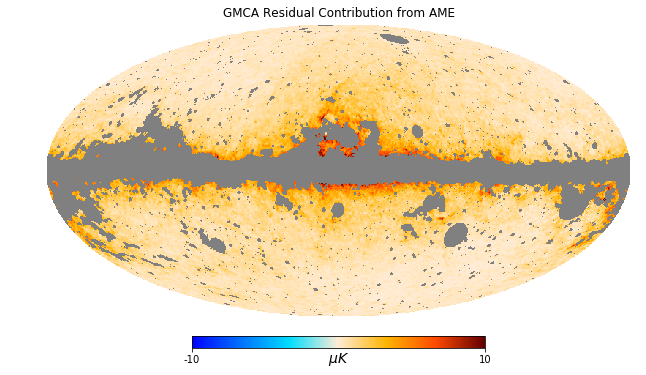}
    \includegraphics[scale=0.35]{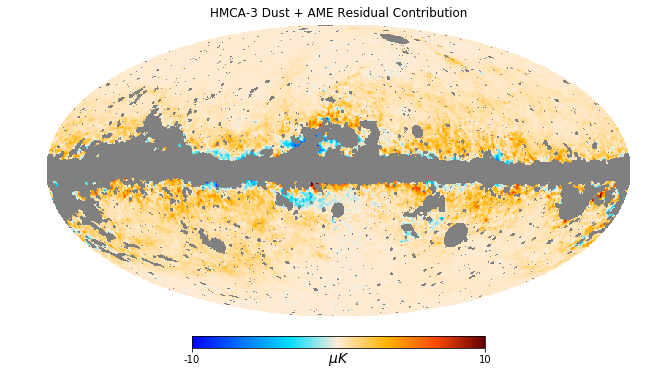}
    \includegraphics[scale=0.35]{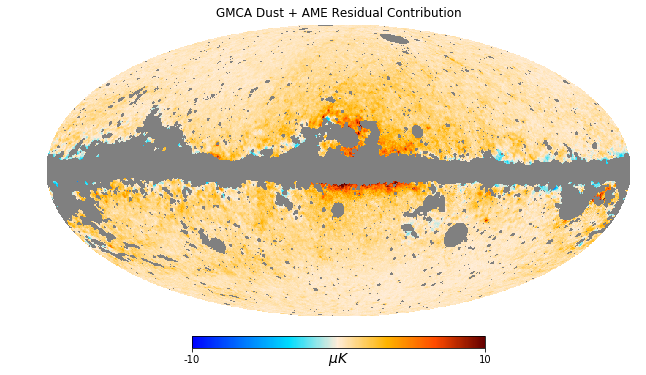}
    \caption{Residual and sum of residual errors from dust and AME in the CMB reconstructions of GMCA and HGMCA-3. Most contamination for GMCA comes from AME (as seen in the middle right-hand image). While HGMCA-3 initially seems to have high residual error in Dust and AME, these contributions are due to a degeneracy in the blind sources separation of correlated sources. This is evidenced by the cancellation of dust and AME errors (as seen in the lower left-hand image), leading to a more accurate reconstruction than GMCA provides.}\label{fig:prop_res}
\end{figure*}

As an alternative to the cross-correlation coefficients we presented in Section \ref{sec:results}, we can also directly propagate through the contributions from the different foreground components to our final CMB fit. If we have a solution for the mixing matrix $A_p$ on patch $p$, then the final source matrix will be set by:
\begin{align}
    S_p &= \left[ A_p \right]_\text{pinv} X,
\end{align}
where $\left[ A_p \right]_\text{pinv}$ is the pseudo-inverse of $A_p$. Since the CMB and foreground components add linearly to $X$, we can rewrite this as: 
\begin{align}
    S_p &= \left[ A_p \right]_\text{pinv} \left( X_\text{CMB} + \sum_\text{foreground} X_\text{foreground} \right).
\end{align}
In a simulation where we have access to each $X_\text{foreground}$, we can isolate the contribution to each source from a specific foreground simply by taking:
\begin{align}
    S_{p,\text{foreground}} &= \left[ A_p \right]_\text{pinv} X_\text{foreground} .
\end{align}
Examining the Dust and AME propagated residuals reveals an interesting degeneracy in the source fitting. In Figure \ref{fig:prop_res}, we show the propagated AME and Dust residuals for HGMCA-3 and GMCA. Individually, both algorithms show significant residual contributions from dust and AME. However, the HGMCA-3 residuals are anti-correlated -  that is they have similar structure but opposite sign. When we add the residuals for both maps, we see that the HGMCA-3 sum cancels most of the large magnitude residuals whereas GMCA does not. Since both the dust and AME maps have similar footprints (they both roughly trace our galaxy) adding a bit more dust and subtracting a bit more AME leads to a similar CMB reconstruction. Thus any algorithm which is blind to these sources, i.e. includes no prior knowledge of frequency dependence, is susceptible to this degeneracy. While this does not create an issue for reconstructing the CMB, such a degeneracy does harm the reconstruction of foreground sources. It would be of interest in future work to explore how incorporating weak foreground priors could help break this degeneracy in the fitting to return more accurate reconstructions of sources besides the CMB.

\label{lastpage}
\end{document}